\documentclass[acmsmall, screen, nonacm]{acmart}
\AtBeginDocument{%
  }
\newif\ifEditMode

\EditModetrue

\usepackage{aliases}
\usepackage{preamble}

\usepackage{enumitem}

\usepackage{booktabs}
\usepackage{longtable}

\usepackage{svg}
\usepackage{pifont}

\begin{document}

\title[There Will Be Spam: Characterizing State-Invariant Transactions and Speculative MEV]{There Will Be Spam: Characterizing State-Invariant Transactions and Speculative MEV}

\author[V. Pahari]{Vabuk Pahari}
\affiliation{%
  \institution{Max Planck Institute for Software Systems (MPI-SWS)}
  \country{Germany}
}
\orcid{0009-0005-1272-0667}

\author[J. Messias ]{Johnnatan Messias}
\affiliation{%
  \institution{Max Planck Institute for Software Systems (MPI-SWS)}
  \country{Germany}
}
\orcid{0000-0002-6021-8402}
\author[C. Ferreira Torres]{Christof Ferreira Torres}
\affiliation{%
  \institution{INESC-ID, Instituto Superior Técnico, University of Lisbon}
  \country{Portugal}
}
\orcid{0000-0001-6992-703X}

\renewcommand{\point}[1]{\par\smallskip\noindent\textbf{#1.} }

\begin{abstract}
Blockchains rely on transparency and immutability to ensure trust, but these guarantees come at the cost of an ever-growing ledger that increasingly threatens decentralization by making it more expensive to store and maintain the full transaction history. In this work, we introduce state-invariant transactions, defined as transactions whose inclusion or removal does not affect the resulting blockchain state beyond transaction fees. We argue that these transactions constitute a form of on-chain spam because they consume execution, bandwidth, storage, and blockspace without contributing to the final ledger state.

We present the first large-scale measurement of state-invariant transactions across Ethereum, Optimism, and Base, identifying nearly 1.4 billion such transactions. While only 2.6\% of Ethereum transactions are state-invariant, they account for 24\% of transactions on Optimism and 37\% on Base, representing a significant source of unnecessary resource consumption on Layer-2 blockchains. We show that speculative Maximal Extractable Value (MEV) is the dominant source of state-invariant transactions on Optimism and Base, accounting for 57\% and 68\%, respectively, but is not the only source as previously assumed. Moreover, despite its popularity, speculative MEV is not the most profitable strategy once the costs of state-invariant transactions are considered. Beyond MEV, we identify substantial malicious activity, with address poisoning campaigns accounting for 53\% of non-reverted state-invariant transactions on Ethereum. Our findings suggest that mitigating state-invariant transactions could substantially reduce blockchain resource consumption and transaction costs while limiting phishing campaigns and other forms of blockchain abuse.
\end{abstract}

\keywords{Blockchain, Decentralized Finance, Speculative, Maximal Extractable Value, Address Poisoning, Inscriptions, Spam}

\maketitle

%------------------------------------------------------------------------------
\section{Introduction}

The advent of blockchain technology has significantly disrupted the traditional financial landscape, giving rise to decentralized finance (DeFi)—a rapidly growing ecosystem that offers open, permissionless, and programmable financial services. Compared to traditional finance (TradFi), DeFi provides several compelling advantages, including enhanced transparency, reduced reliance on intermediaries, and global accessibility. These benefits have fueled explosive growth: the total value locked (TVL) across DeFi protocols has reached hundreds of billions of dollars \cite{defillama}, with a continuous influx of users and the steady deployment of new protocols and applications.

However, the rise of DeFi has also ushered in a new era of predatory trading practices. 
These practices have taken on a novel dimension within DeFi by exploiting the structural and technological nuances of decentralized systems.
This phenomenon is commonly referred to as maximal extractable value (MEV), which denotes the profit that can be gained by reordering, inserting, or censoring transactions within a block. 
Unlike traditional finance, where regulation limits the scope of such manipulation, DeFi remains largely unregulated, creating a "wild west" environment where MEV strategies flourish unchecked.

A frequently cited root cause of negative MEV is the existence of a public mempool, which exposes pending user transactions to adversaries. 
Rollups, a new class of Layer 2 scaling solutions built on top of Ethereum, address this concern by implementing centralized transaction ordering and private mempools. 
This architectural shift not only theoretically mitigates exposure to harmful MEV but also significantly lowers transaction fees through batching and compression mechanisms. 
The Dencun upgrade, introduced in March 2024 \cite{dencun}, further amplified these advantages by enabling blob transactions, which drastically reduced rollup fees. 
Additionally, rollups provide significantly shorter block times than Ethereum, with blocks produced approximately every 2 seconds on Optimism and Base compared to every 12 seconds on Ethereum, enabling faster transaction inclusion.

While these developments appear beneficial for both users and the ecosystem, an emerging challenge has begun to surface: the rise of \textit{speculative MEV}.
In an environment without a public mempool, traditional MEV search techniques based on pure off-chain computations are no longer effective.
Instead, MEV participants are increasingly relying on on-chain searching to speculatively insert transactions, attempting to capture MEV opportunities such as arbitrage and liquidations, without full knowledge of pending transactions.
This speculative approach introduces a new dimension of side-effects such as computational spam that has yet to be systematically studied.

Existing literature \cite{solmaz@AFT25,wang2026blockspace,wu2026wait} has primarily focused on measuring spam across blockchains assuming its sole source is related to speculative MEV originating from arbitrages.
In contrast, this work is the first to formally define a metric for quantifying computational spam through the notion of \emph{state-invariant transactions}. We define state-invariant transactions as transactions that consume blockchain resources without producing any meaningful state changes beyond transaction fees, such that removing them from the blockchain leaves the resulting ledger state unchanged.

In rollups, where block times are extremely short, off-chain MEV searches are impractical due to latency constraints. Instead, searchers submit speculative transactions on-chain, betting on the potential to extract value.
While economically viable due to low transaction costs, these speculative transactions often fail to yield any profit and consume valuable gas and blockspace, ultimately introducing computational spam into the system.

This practice not only degrades user experience by delaying the inclusion of legitimate transactions but also incentivizes arms races. 
On priority-fee-based rollups like Optimism and Base, speculative MEV activity fuels priority gas auctions (PGAs), thereby raising transaction costs for all users. 
Moreover, the inclusion of functionally meaningless speculative transactions increases blockchain bloat and creates potential vectors for censorship and denial-of-service (DoS) attacks as running full archive nodes becomes more of a challenge.

In this paper, we conduct the first formal study of spam through our definition of state-invariant transactions across Ethereum, Optimism, and Base. Our large-scale measurement across Ethereum, Optimism, and Base identifies nearly 1.4 billion state-invariant transactions. Although they comprise just 2.6\% of transactions on Ethereum, they represent 24\% and 37\% of all transactions on Optimism and Base, respectively, revealing a major source of unnecessary execution, storage, and blockspace consumption on Layer-2 networks. We find that speculative MEV is indeed the primary driver of state-invariant transactions on Optimism and Base, accounting for 57\% and 68\%, respectively, but is far from being the only source. Interestingly, when transaction fees incurred by state-invariant transactions are taken into account, speculative MEV becomes unprofitable on average. We further uncover a substantial volume of state-invariant transactions arising from malicious activity asscociated with address poisoning campaigns accounting for 53\% of all non-reverted state-invariant transactions on Ethereum. These findings demonstrate that mitigating state-invariant transactions could significantly reduce blockchain resource consumption while simultaneously limiting phishing campaigns and other forms of on-chain abuse.

\noindent
\textbf{Contributions.} Our main contributions are:

\begin{itemize}[leftmargin=*]

\item We introduce the notion of state-invariant transactions, a novel, systematic method for measuring and comparing spam across EVM-compatible blockchains, and present the first large-scale study of such transactions on Ethereum, Base, and Optimism from September 1, 2023 to July 31, 2025.

\item We identify nearly 1.4 billion state-invariant transactions. While they comprise only 2.6\% of all transactions on Ethereum, they represent 24\% and 37\% of all transactions on Optimism
and Base, respectively, highlighting a significant reduction in storage space if removed from the history.

\item We show that speculative MEV extends beyond mere arbitrage and also includes speculative liquidations. We develop a scalable methodology for classifying MEV bots as speculative, non-speculative, or hybrid. We also find that most speculative arbitrage bots on Optimism and Base turn out to be unprofitable when costs related to state-invariant transactions are added.

\item We demonstrate that, although speculative MEV is a major source of state-invariant transactions, 57\% and 68\%, on Optimism  and Base, respectively, a substantial fraction of spam arises from other activities, including reverting transactions (e.g., due to slippage protection), phishing (e.g., address poisoning) and on-chain message passing (e.g., inscriptions). On Ethereum, 53\% of non-reverting state-invariant transactions are associated to address poisoning.

\end{itemize}

\section{Background}\label{sec:background}

This section provides the necessary background on Ethereum, Layer-2 blockchains, Maximal Extractable Value (MEV), address poisoning, and inscriptions.

\subsection{Ethereum and Layer-2 Blockchains}

Ethereum is a blockchain platform that enables the transfer of cryptocurrencies between accounts and the execution of smart contracts (i.e., programs) through transactions. Transactions are grouped into sequentially linked blocks, forming an immutable ledger in which each block references its predecessor. Table~\ref{tab:blockchain_comparison} summarizes the key characteristics of the blockchains considered in this work.

Ethereum is a Layer-1 (L1) blockchain that relies on a decentralized consensus protocol for block production. While this design provides strong security and decentralization guarantees, it also limits throughput due to a block time of approximately 12 seconds and a constrained block gas limit, resulting in relatively high transaction fees during periods of congestion. To address these limitations, Layer-2 (L2) scaling solutions, such as Optimism and Base have been proposed, which execute transactions off-chain and periodically commit their state to Ethereum. Instead of decentralized consensus, these rollups rely on a centralized sequencer that is responsible for transaction ordering and block production, enabling significantly shorter block times of approximately 2 seconds, higher throughput, and lower transaction fees.

\begin{table}[b]
    \centering
    \begin{adjustbox}{max width=\linewidth}
    \begin{tabular}{l c c c l r c c r}
        \toprule
         & & \textbf{Public} & \textbf{Bundling} & \multicolumn{1}{c}{\textbf{Transaction}} & \multicolumn{1}{c}{\textbf{Block}} & \textbf{Transaction} & & \multicolumn{1}{c}{\textbf{Archive}} \\
        \textbf{Chain} & \textbf{Layer} & \textbf{Mempool} & \textbf{Service} & \multicolumn{1}{c}{\textbf{Ordering}} & \multicolumn{1}{c}{\textbf{Time}} & \textbf{Fees} & \textbf{Throughput} & \multicolumn{1}{c}{\textbf{Node Size}} \\
        \midrule
        Ethereum & L1 & Yes & Yes & Gas Price & 12s & High & Low &  3.5 TB \\
        Optimism & L2 & No & No & Gas Price & 2s & Low & High & 3.9 TB \\
        Base & L2 & No & No & Gas Price+FCFS & 2s & Low & High & 9.0 TB \\
        \bottomrule
    \end{tabular}
    \end{adjustbox}
    \caption{Comparison of key L1 and L2 blockchain characteristics across Ethereum, Optimism, and Base.}
    \label{tab:blockchain_comparison}
\end{table}

Unlike Ethereum, which exposes pending transactions through a public mempool, optimistic rollups typically submit transactions directly to the sequencer. Consequently, pending transactions are generally not publicly observable prior to inclusion in a block (i.e., private mempool). Nonetheless, the Ethereum ecosystem also supports private mempools that provide transaction bundling services with revert protection and atomic execution guarantees. Under this model, bundles (i.e., groups of transactions) are forwarded for inclusion only if all constituent transactions execute successfully and offer a competitive inclusion fee. Comparable bundling infrastructure is currently absent on L2s.
Across the studied chains, transactions are primarily ordered according to their offered transaction fees (i.e., gas price), while ties are typically resolved on a first-come, first-served basis on Base.
Because blockchains permanently record all transactions, their storage requirements continuously increase over time. Although L2 rollups substantially improve throughput, they also generate data at a much higher rate than Ethereum, resulting in faster blockchain growth and increasing the storage burden on nodes that maintain a local copy of the chain for independent verification. Table~\ref{tab:blockchain_comparison} reports the storage consumption of a Reth \cite{tracing} archive node across the blockchains considered in this work. As shown in Table~\ref{tab:blockchain_comparison}, the complete history of Ethereum currently occupies approximately 3.5 TB, compared to 3.9 TB for Optimism and 9.0 TB for Base. This difference is particularly striking given that Ethereum is already operating since 2015, whereas Optimism and Base were introduced only in 2021 and 2023, respectively.

\subsection{Maximal Extractable Value}

The growth of decentralized finance (DeFi) has led to the widespread emergence of Maximal Extractable Value (MEV), which refers to the profit obtained by strategically influencing the ordering of transactions within a block. MEV is typically extracted by specialized participants, known as searchers, that operate automated bots to identify profitable opportunities. On Ethereum, searchers monitor pending transactions in the public mempool, whereas on rollups they primarily analyze recently produced blocks due to the absence of a publicly accessible mempool.

The three most common MEV strategies are arbitrage, liquidation, and sandwiching. Arbitrage exploits temporary price discrepancies across decentralized exchanges by purchasing an asset on one exchange and simultaneously selling it on another at a higher price. Liquidation targets undercollateralized lending positions, where searchers repay part of a borrower's debt in exchange for acquiring collateral at a discount. Such opportunities are often triggered by price oracle updates that change the valuation of deposited collateral. Sandwiching exploits the visibility of pending user transactions by placing transactions immediately before and after a users pending trade. The sandwicher first purchases the asset before the user's transaction, causing the user to execute at a less favorable price, and subsequently sells the acquired asset after the user's trade at a profit.

Recent studies~\cite{solmaz@AFT25,wu2026wait} report a substantial increase in transaction volume on Layer-2 (L2) blockchains, much of which is attributed to speculative MEV. Unlike Ethereum, optimistic rollups do not expose a public mempool and charge comparatively low transaction fees. Consequently, rather than identifying profitable opportunities off-chain before submitting transactions, searchers increasingly speculate on MEV opportunities by submitting transactions that probe for arbitrage or liquidation opportunities during execution. This strategy can provide a competitive advantage over searchers that rely solely on information from previously finalized blocks.

However, speculative MEV requires searchers to issue numerous probing transactions, the vast majority of which do not result in a successful extraction. Although these transactions consume block space and permanently increase the blockchain's history, they typically leave no meaningful state changes. On Ethereum, speculative execution is largely mitigated through transaction bundling services, which simulate transactions off-chain and forward them for inclusion only if they successfully extract MEV. Comparable infrastructure is generally unavailable on L2s. Instead, low transaction fees makes on-chain speculation economically viable despite frequent failed attempts. Moreover, the short block intervals leave searchers with limited time for off-chain opportunity discovery, further incentivizing them to perform the search process directly on-chain.

\subsection{Address Poisoning}

Address poisoning is a phishing attack that exploits the abbreviated display of blockchain addresses in cryptocurrency wallets, where users often rely only on the address prefix and suffix for identification. Attackers generate look-alike addresses with matching prefixes and suffixes and monitor the blockchain for transactions involving addresses that resemble those they control. After identifying a suitable transaction, the attacker sends a phishing transaction from the look-alike address to one of the transaction participants. Because both the legitimate and phishing transactions appear in the victim's transaction history, the victim may mistakenly copy the attacker's address when initiating a future transfer, resulting in the loss of funds.

Address poisoning attacks can be broadly classified into ETH-based and token-based attacks. Regardless of the asset used, attackers typically employ one of the three techniques: \emph{zero-value}, \emph{dust-value}, and \emph{fake-token} transfers \cite{tsuchiya2025blockchainaddresspoisoning,abs-2508-12107}. 
Zero-value transfers send no assets (i.e., a transfer amount of zero) to the victim. As a result, neither ETH nor token balances are modified, while a `Transfer` event is still emitted, creating a misleading transaction history entry at minimal cost.
Dust-value transfers send a negligible amount of ETH or tokens to increase the legitimacy of the phishing transaction. 
Fake-token transfers use attacker-deployed ERC-20 tokens to mimic the transfer amounts of legitimate transactions, making phishing transactions appear nearly indistinguishable from genuine transfers without requiring the attacker to spend valuable assets.

\subsection{On-Chain Postings and Inscriptions}

On-chain posting is the process of embedding data or text directly into a blockchain transaction. This enables users to store and disseminate data in a censorship-resistant manner by permanently embedding arbitrary structured data into transaction calldata.
Inspired by Bitcoin Ordinals~\cite{Wang-Ordinals-25,BTC-Ordinals} and the BRC-20 protocol~\cite{Wang-Ordinals-25}, inscriptions were later adapted to Ethereum and other EVM-compatible blockchains~\cite{messiasinscription}, where users encode JSON-formatted messages describing operations such as `deploy', `mint', and `transfer'. Unlike ERC-20 tokens or NFTs, inscriptions do not rely on smart contracts. Instead, off-chain indexers reconstruct token ownership and balances by interpreting the embedded messages stored on-chain.

Inscription transactions are identified by inspecting the transactions input calldata. 
The calldata encodes a JSON payload containing fields such as inscription protocol, operation, token identifier, and token amount, which can be decoded to reconstruct the inscription medatada~\cite{messiasinscription}. The popularity of inscriptions led to a large surge in transactions on blockchains. During peak periods, inscriptions accounted for nearly 90\% of all transactions on some Layer-2 rollups, with over 99\% corresponding to speculative token minting rather than actual trading~\cite{messiasinscription}. 
Although they consume blockspace and transaction fees, the blockchain itself does not interpret the embedded payload or execute application-specific logic. Instead, the calldata merely serves as a communication channel for off-chain indexers, which reconstruct the inscription state independently of the blockchain.

\section{Methodology}
\label{sec:methodology}

In this section, we formalize the notion of state-invariant transactions and present our methodology for measuring and classifying them at scale.

\subsection{Detecting State-Invariant Transactions}

While prior work~\cite{solmaz@AFT25,wu2026wait} has investigated ``spam'' transactions, their identification largely relies on first detecting successful arbitrage transactions of MEV bots and subsequently classifying transactions that either revert or do not execute trades as spam.
However, this methodology assumes that spam transactions are exclusively associated with arbitrage, potentially underestimating the true prevalence of spam transactions associated to other types of MEV such as liquidations. Moreover, by focusing solely on MEV-related activity, it fails to capture spam transactions potentially arising from other behaviors.

We take a more general approach by identifying the broader class of state-invariant transactions.
We define a state-invariant transaction as a transaction that executes EVM code but does not modify the blockchain state with respect to account storage or account balances, except for changes resulting from transaction fees.
Although such transactions are included on-chain, their execution has no effect on the final state beyond the transaction fees paid by the sender.
Equivalently, if a state-invariant transaction had not been included in a block, the resulting blockchain state would be identical.
Note that this definition excludes simple native token transfers between externally owned accounts (EOA), as these transactions do not execute any EVM code.

Formally, let ${a_1, a_2, \ldots, a_n}$ denote the set of accounts in the blockchain. Each account $a_i$ is represented as the tuple $a_i = (b_i, s_i)$, where $b_i$ denotes the account balance and $s_i$ denotes its storage.
The storage of an account is only relevant for smart contract accounts. Hence, for any EOA $a_i$, $s_i = \emptyset$.
Define a transaction, $tx$ as $tx = \{tx_{from}, tx_{to}, tx_{input}, tx_{value} \}$.
If $tx_{to}$ is not a smart contract address, then $tx_{input} = \emptyset$.
We define the blockchain state as the set of all accounts, denoted by $S = {a_1, a_2, \ldots, a_n}$. Consequently, the blockchain state captures the balance and storage associated with every account.
Let $S_{B_i}$ denote the state of the blockchain after the execution of all transactions through block $B_i$.
Given a block $B_{i+1}$ with transaction sequence $[tx_1, tx_2, \ldots, tx_n]$, let $S_{B_i, tx_j}$ denote the blockchain state obtained by starting from $S_{B_i}$ and executing the transactions $tx_1, tx_2, \ldots, tx_j$ in order.
Consequently, the final state after executing block $B_{i+1}$ is given by $S_{B_{i+1}} = S_{B_i, tx_n}$.

We define a transaction $t_s$ to be state-invariant if its removal from a block does not affect the resulting blockchain state, excluding the transfer of transaction fees to the block proposer. Formally, consider a block
$B_i = [tx_1, tx_2, \ldots, t_{s-1}, t_s, t_{s+1}, \ldots, tx_n]$
and the corresponding block
$B'i = [tx_1, tx_2, \ldots, t_{s-1}, t_{s+1}, \ldots, tx_n],$
obtained by removing $t_s$. Then, $t_s$ is state-invariant if
$S_{B_i} = S_{B'_i}$,
where the comparison ignores changes to the block proposer's balance resulting from the transaction fees paid by $t_s$.
Intuitively, a transaction that invokes a smart contract but leaves the blockchain state unchanged, apart from the payment of transaction fees, is classified as a state-invariant transaction.
Under this definition, all reverted smart contract transactions are considered state-invariant, since their execution does not modify the blockchain state beyond transferring transaction fees.

Following our definition, state-invariant transactions could be identified by iteratively removing each transaction from a block, replaying the modified block, and comparing its resulting state to that of the original block. If both states are identical, the removed transaction is classified as state-invariant. Repeating this procedure for every transaction in every block, however, would incur substantial computational overhead and is therefore impractical at scale.
Instead, to identify state-invariant transactions more efficiently, we rely on heuristics and leverage the \texttt{tracing.trace\_replay\_block\_transactions} function provided by the Web3 library \cite{tracing}, which returns the ``stateDiff,'' i.e., the change in state corresponding to each account on the blockchain, associated with each transaction per block. Because this approach traces the state difference only once per block while analyzing multiple transactions within that block, it is significantly more efficient than tracing each transaction individually.

For Ethereum, we identify transactions whose only state changes involve the transaction sender and the block proposer (i.e., ``miner'' field) in the block.
These changes account solely for the payment of transaction fees. 
We further verify that the observed state changes exactly match the transaction fees paid. The sender's balance must decrease by the total transaction fee paid, while the block proposer's balance must increase by the total fee excluding the portion of the base fee that is burned.

For layer-2 rollups such as Optimism and Base, transaction fees are distributed across multiple fee vaults rather than being paid directly to a single block proposer. 
Specifically, each transaction fee is split into three components: the base fee, the sequencer fee (typically corresponding to the priority fee), and the L1 fee, which compensates for posting transaction data to layer-1 (i.e., Ethereum). These components are deposited into their respective fee vaults.
Accordingly, our methodology for Optimism and Base mirrors our methodology used for Ethereum. 
We verify that the only state changes involve the transaction sender and the corresponding fee vaults, and that the balance changes exactly match the expected base fee, sequencer fee, and L1 fee associated with the transaction. 
Transactions that modify the state of any additional account are not considered state-invariant under our definition.

\subsection{Classifying State-Invariant Transactions}

Having identified state-invariant transactions, we observe that they arise for a variety of reasons. To better understand these underlying causes, we develop a set of heuristics to classify these transactions.

\subsubsection{Identifying Speculative MEV}

Previous work \cite{solmaz@AFT25,wu2026wait} has leveraged trace-level heuristics to classify arbitrage transactions or bots as either speculative or non-speculative. Such approaches obtain execution traces from a full archive node and analyze characteristics such as the number of \texttt{STATICCALL}s performed, the contract addresses accessed, and whether specific read or simulation functions are invoked. These heuristics are then used to infer whether the search process was conducted primarily off-chain or on-chain. However, trace-based methodologies are computationally expensive and do not scale well. Consequently, prior studies often analyze only a subset of transactions \cite{wu2026wait}, rather than entire datasets, to draw conclusions. This sampling-based approach may overlook bots that employ hybrid search strategies.

Our approach follows a trace-independent methodology based on a simple observation. We begin by applying state-of-the-art arbitrage and liquidation detection techniques by Torres et al.~\cite{Torres-rollingintheshadows-CCS} to construct a dataset of successful arbitrage and liquidation transactions. The dataset provides high-level transaction metadata, including pool addresses, token addresses, and the corresponding token amounts involved in each arbitrage and liquidation.
The key insight of our methodology is that both speculative and non-speculative searchers can potentially hard-code arbitrage and liquidation paths directly into their smart contracts. Such paths consist primarily of pool and token addresses and are frequently embedded within the contract for gas optimization purposes. Rather than dynamically supplying this information via calldata, searchers can deploy a new contract whenever paths or supported assets need to be added or removed. This design reduces calldata size and transaction costs, regardless of whether the underlying search process is speculative or non-speculative, making such information unsuitable for distinguishing between the two.

\begin{figure*}[t!]
    \centering
    \begin{subfigure}[t]{0.5\textwidth}
        \centering
        \includegraphics[height=1.1in]{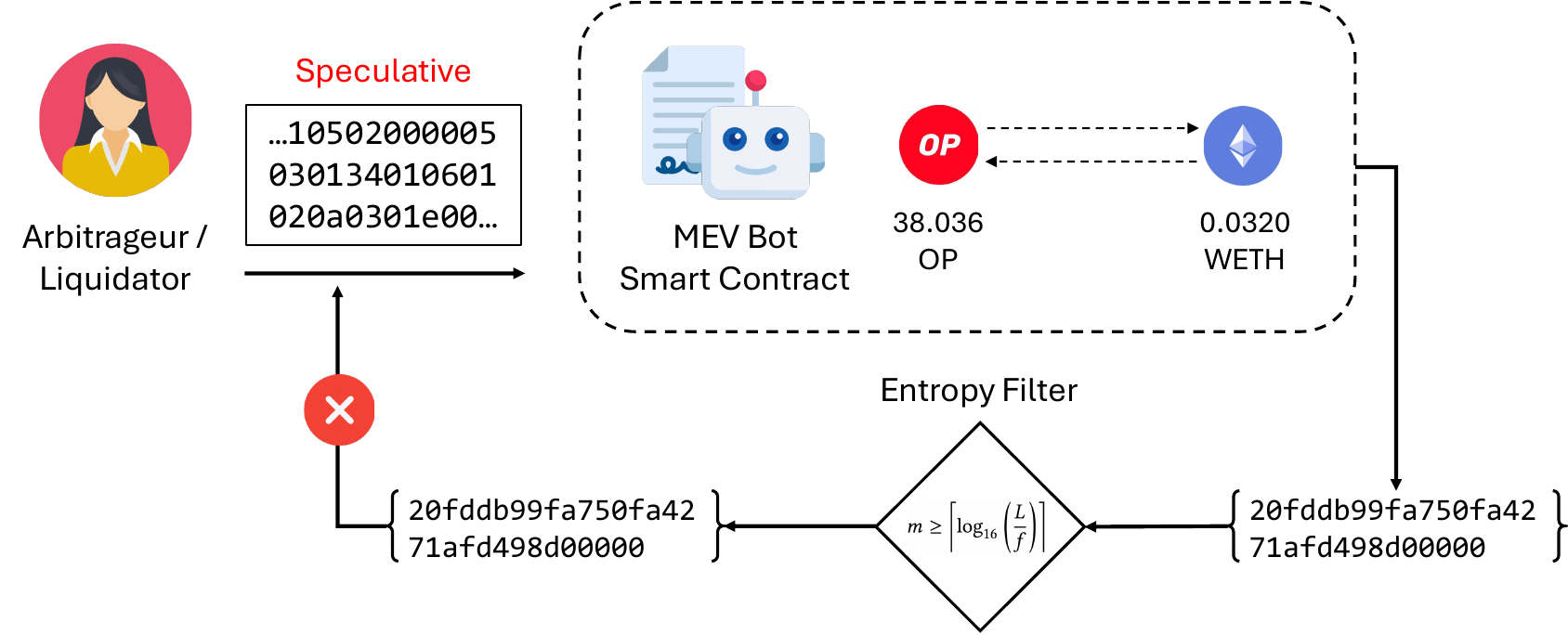}
        \caption{Speculative MEV Example}
    \end{subfigure}%
    ~ 
    \begin{subfigure}[t]{0.5\textwidth}
        \centering
        \includegraphics[height=1.1in]{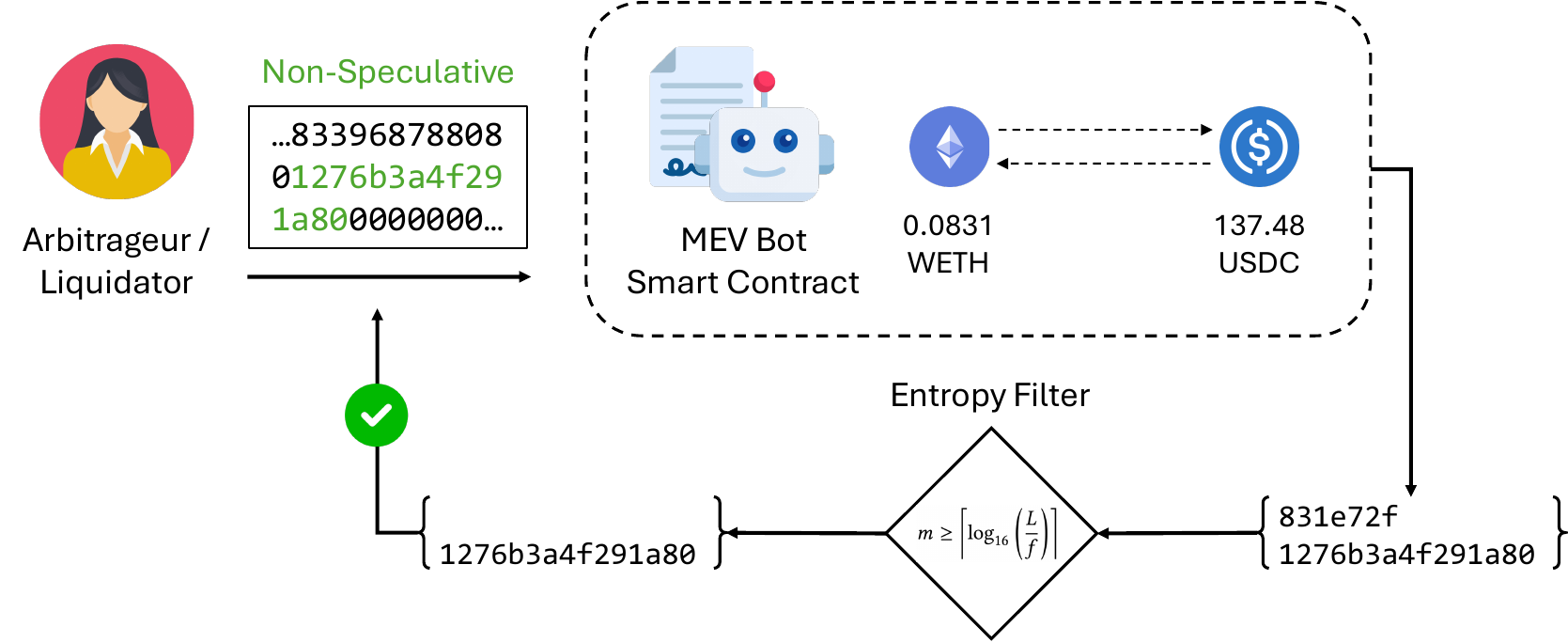}
        \caption{Non-Speculative MEV Example}
    \end{subfigure}
    \caption{Our methodology for distinguishing between speculative and non-speculative MEV is based on the observation that when bot operators precompute token amounts off-chain, these values must be explicitly encoded in the transaction calldata. Consequently, transactions containing such precomputed values are classified as non-speculative, since bot operators are not speculating about an MEV opportunity as they have identified it already off-chain.}
    \label{fig:speculative_detection_methodology}
\end{figure*}

In contrast, token amounts are inherently dynamic because they depend on the current state of the underlying liquidity pools and market conditions. Consequently, token amounts cannot generally be predetermined and hard-coded into a contract. We therefore argue that the presence of token amount information within calldata provides strong evidence that these values were computed off-chain prior to transaction submission. In other words, the searcher has already performed part of the search process off-chain and is therefore engaging in a non-speculative strategy.
\figureautorefname{} \ref{fig:speculative_detection_methodology} provides a high-level overview of our feedback-loop-based labeling methodology. For each detected successful arbitrage or liquidation, we extract all token amounts involved in the transaction and check whether at least one of these values appears in the transaction calldata. This procedure enables classification of transactions within milliseconds and does not require computationally expensive tracing.

A challenge of this approach is the possibility of coincidental matches between token amounts and arbitrary calldata subsequences. Ideally, calldata would be parsed according to the contract's Application Binary Interface (ABI), allowing individual parameters to be analyzed directly. However, because MEV bot source code is typically proprietary, ABIs are not available. Although existing reverse-engineering tools can attempt to recover ABIs from bytecode using symbolic execution and data-flow analysis \cite{evmmole,whatsabi}, these techniques are imperfect and frequently fail for contracts employing custom calldata parsing logic, particularly when implemented within fallback functions.
To mitigate the likelihood of accidental matches, we exploit the entropy of the hexadecimal representation of token amounts. Specifically, we compute the minimum hexadecimal string length that a token amount must have before it is considered eligible for matching against calldata:
$$
m \ge \left\lceil \log_{16}\left(\frac{L}{f}\right) \right\rceil
$$
where $m$ denotes the hexadecimal string length of the token amount, $L$ denotes the hexadecimal string length of the calldata, and $f$ represents the tolerated false-positive rate. In our evaluation, we set $f = 10^{-6}$, corresponding to a maximum false-positive probability of one in one million. Consequently, only token amounts whose hexadecimal representations satisfy the minimum required length are considered during matching.
After applying this filtering step, a transaction is labeled as non-speculative if at least one eligible token amount is found within the calldata. Otherwise, the transaction is labeled as speculative. We perform this classification for every successful arbitrage and liquidation transaction associated with a given bot and subsequently classify bots into three categories: 
(a) \emph{speculative:} all observed MEV transactions are labeled as speculative;
(b) \emph{non-speculative:} all observed MEV transactions are labeled as non-speculative;
(c) \emph{hybrid:} both speculative and non-speculative MEV transactions are observed.

Hybrid MEV bots employ a variety of strategies. We observed that some bots initially operated using a non-speculative strategy before their bytecode was later replaced to implement speculative execution. In other cases, the deployed bytecode supported both strategies, directly allowing searchers to switch between them over time. Most commonly, however, hybrid bots compute a minimum input token amount off-chain and include it in the submitted calldata. During execution, the contract performs an additional on-chain computation to determine whether market conditions have changed. If the optimal input amount has increased, the larger value is used; if it has decreased, the arbitrage is aborted. By combining off-chain pre-computation with on-chain decision making, these bots exemplify a hybrid search strategy.

\subsubsection{Identifying Address Poisoning}

We identify two categories of state-invariant transactions associated with address poisoning attacks.
The first category consists of zero-value transfers, in which no tokens are actually transferred between accounts, yet a `Transfer' event is emitted. Since no token balances are modified, these transactions are state-invariant.
The second category comprises transactions involving `fake' or `tampered' ERC-20 token contracts. 
These contracts emit `Transfer' events indicating that tokens have been transferred, but these tokens are implemented such that the on-chain state is never actually modified. A likely motivation for this design is to avoid the storage writes required by a genuine token transfer, thereby reducing the gas cost of each address poisoning transaction since wallets solely rely on emitted events to construct token transfer history.
However, previous works \cite{tsuchiya2025blockchainaddresspoisoning,abs-2508-12107} related to address poisoning also introduces so-called `dust' attacks, in which an attacker transfers a very small amount of a legitimate ERC-20 token (e.g., USDC or USDT) to a victim in an attempt to deceive them.
Such attacks are intentionally excluded from our methodology because they involve genuine state changes and token spendings for attackers and therefore do not satisfy our definition of a state-invariant transaction.
Moreover, previous work \cite{tsuchiya2025blockchainaddresspoisoning} shows that `dust' attacks only account for 1.7\% of address poisoning attacks in the wild.
We label state-invariant transactions as address poisoning transactions using the public dataset by Tsuchiya \ea~\cite{tsuchiya2025blockchainaddresspoisoning} and distinguish between zero-value transfers and fake-token transfers by checking whether the transferred amount is zero or not.

\subsubsection{Identifying On-Chain Postings and Inscriptions}
To identify inscription transactions, we follow the methodology proposed by Messias et al.~\cite{messiasinscription}. Specifically, we first identify self-transactions, where the transaction sender address is equivalent to the destination address of the transactions (i.e., \texttt{from} $==$ \texttt{to}) and then inspect their calldata, classifying a transaction as an inscription if its calldata starts with the hexadecimal prefix \texttt{0x646174613a}, corresponding to the ASCII string `\texttt{data:}'.

\subsection{Limitations}

Our proposed methodology to classify state-invariant transactions has ultimately some limitations.
For instance, our classification of speculative and non-speculative MEV is not robust to searchers obfuscating or encrypting transaction calldata. In such cases, token amounts can no longer be reliably extracted from the calldata, potentially causing non-speculative MEV bots to be misclassified as speculative. However, obfuscating or encrypting calldata incurs additional computational and implementation costs, which may limit the prevalence of such techniques in practice. Another limitation is the possibility of substring collisions, whereby a token amount coincidentally matches an unrelated calldata substring. To mitigate this issue, we adopt a conservative false-positive probability threshold of $10^{-6}$, making such collisions exceedingly unlikely.

To identify address poisoning attacks, we rely on the dataset of Tsuchiya et al.~\cite{Tsuchiya-Amplification-Attack-25}. However, the publicly available dataset covers Ethereum only up to block 20,208,026 and does not include labels for Base or Optimism. Although the authors kindly provided an extended dataset covering Ethereum blocks 21,565,106 to 23,870,337, our final dataset still contains a gap between blocks 20,208,026 and 21,565,106.

\section{Empirical Analysis}

In this section, we present the results and insights of our large-scale measurement study of state-invariant transactions.

\begin{table}
    \centering
    \begin{adjustbox}{max width=\columnwidth}
    \begin{tabular}{l r r r r r r r r}
       \toprule
       & & & \multicolumn{3}{c}{\textbf{Arbitrage}} & \multicolumn{3}{c}{\textbf{Liquidation}} \\
       \cline{4-9} \noalign{\vskip 3pt}
       \textbf{Chain} & \multicolumn{1}{c}{\textbf{Block Range}} & \multicolumn{1}{c}{\textbf{Blocks}} & \textbf{Transactions} & \textbf{Bots} & \textbf{Pools} & \textbf{Transactions} & \textbf{Bots} & \textbf{Pools} \\
       \midrule
       Ethereum & 18,037,988 - 23,042,513 & 5,004,526 & 1,680,903 & 1,680 & 24,288 & 31,624 & 256 & 226 \\
       Base  & 3,368,527 - 33,608,526 & 30,240,000 & 14,443,440 & 10,087 & 2,324 & 38,461 & 688 & 65 \\
       Optimism & 108,963,812 - 139,203,811 & 30,240,000 & 4,339,358 & 2,495 & 33,985 & 24,703 & 511  & 82\\
       \midrule
       \textbf{Total} & & & \textbf{20,463,701} & \textbf{14,262} & \textbf{60,597} & \textbf{94,788} & \textbf{1,455} & \textbf{373} \\
       \bottomrule
    \end{tabular}
    \end{adjustbox}
    \caption{Overview of detected arbitrage and liquidation metadata across Ethereum, Optimism, and Base.}
    \label{tab:mev_data_range}
\end{table}

\subsection{Data Collection}

We deployed archive nodes for Ethereum, Optimism, and Base using Reth~\cite{tracing}. Our measurement period spans from September 1, 2023 at 00:00:00, to July 31, 2025 at 23:59:59, comprising 5,004,526 blocks on Ethereum and 30,240,000 blocks each on Optimism and Base. These blocks contain 837,457,443 transactions on Ethereum, 501,969,834 on Optimism, and 3,255,619,515 on Base. We apply our methodology to this dataset to identify and classify state-invariant transactions. Moreover, \tableautorefname{} \ref{tab:mev_data_range} summarizes the number of successful arbitrages and liquidations that we have extracted using our extended version of Torres et al.'s \cite{Torres-rollingintheshadows-CCS} methodology. Overall we have found a total of 20,463,701 arbitrage transactions across 14,262 bots and 60,597 pools as well as 94,788 liquidation transactions across 1,455 bots and 373 pools, combined across Ethereum, Optimism, and Base.

\subsection{Overview of Empirical Results}

Table~\ref{tab:spam-txs} provides an overview of the state-invariant transactions identified by our methodology over the measurement period from September 1, 2023, to July 31, 2025.
We identify 22,830,229 state-invariant transactions on Ethereum, representing approximately 2.6\% of all Ethereum transactions in our dataset.
In contrast, the Layer-2 blockchains Optimism and Base exhibit substantially higher proportions of transactions classified as state-invariant transactions.
Specifically, we identify 122,454,486 state-invariant transactions on Optimism, accounting for approximately 24.4\% of all transactions, while Base contains over 1.2 billion state-invariant transactions, corresponding to 37.5\% of all transactions. 
These findings suggest that although Layer-2 networks significantly increase transaction throughput, a considerable share of this additional activity consists of state-invariant transactions.

\begin{table*}[b]
\begin{center}
\begin{adjustbox}{max width=\linewidth}
\begin{tabular}{l r r r r r r}
\toprule
 & \multicolumn{2}{c}{\textbf{Ethereum}} & \multicolumn{2}{c}{\textbf{Optimism}} & \multicolumn{2}{c}{\textbf{Base}} \\
  \cline{2-7} \noalign{\vskip 3pt}
 \textbf{State-Inv. Category} & \textbf{Transactions} & \multicolumn{1}{c}{\textbf{Earnings [ETH]}} & \textbf{Transactions} & \multicolumn{1}{c}{\textbf{Earnings [ETH]}} & \textbf{Transactions} & \multicolumn{1}{c}{\textbf{Earnings [ETH]}} \\ 
\midrule
\textbf{Failed Transactions} & \textbf{15,783,915 (69.1\%)} & \textbf{4,541.6 (88.3\%)} & \textbf{23,732,136 (19.4\%)} & \textbf{724.2 (80.9\%)} & \textbf{400,704,401 (32.6\%)} & \textbf{4,816.25 (67.7\%)} \\
\hspace{0.2cm} Reverted & 13,110,091 (57.4\%) & 3,939.5 (76.6\%) & 21,392,488 (17.5\%) & 641.4 (71.7\%) & 379,360,903 (30.9\%) & 4,449.75 (62.6\%)  \\
\hspace{0.2cm} Out of Gas & 2,673,824 (11.7\%) & 602.1 (11.7\%) & 2,339,648 (1.9\%) & 82.8 (9.3\%) & 21,343,498 (1.7\%) & 366.5 (5.2\%) \\
\textbf{Successful Transactions} & \textbf{7,046,314 (30.9\%)} & \textbf{600.0 (11.7\%)} & \textbf{98,722,350 (80.6\%)} & \textbf{170.5 (19.1\%)} & \textbf{828,221,745 (67.4\%)} & \textbf{2,294.2 (32.3\%)} \\
\hspace{0.2cm} With Logs & 5,802,659 (25.4\%) & 411.6 (8.0\%) & 8,299,857 (6.8\%) & 55.8 (6.2\%) & 57,335,881 (4.7\%) & 291.4 (4.1\%) \\
\hspace{0.4cm} Transfer Logs & 3,103,840 (13.6\%) & 287.8 (5.6\%) & 2,054,009 (1.7\%) & 16.3 (1.8\%) & 6,372,445 (0.5\%) & 28.2 (0.4\%) \\
\hspace{0.6cm} Zero Transfer & 973,477 (4.3\%) & 164.9 (3.2\%) & 1,177,219 (1.0\%) & 5.3 (0.6\%) & 1,903,424 (0.2\%) & 6.47 (0.1\%) \\
\hspace{0.6cm} Same Transfer & 325,358 (1.4\%) & 13.4 (0.3\%) & 52,416 (0.05\%) & 0.55 (0.1\%) & 248,659 (0.0\%) & 0.98 (0.0\%) \\
\hspace{0.4cm} Same Approval Logs & 2,172,004 (9.5\%) & 77.7 (1.5\%) & 1,418,330 (1.2\%) & 10.43 (1.2\%) & 6,067,637 (0.5\%) & 20.11 (0.3\%) \\
\hspace{0.4cm} Other & 526,815 (2.3\%) & 52.4 (1.0\%) & 4,827,518 (3.9\%) & 29.1 (3.3\%) & 44,895,799 (3.7\%) & 243.1 (3.4\%) \\
\hspace{0.2cm} Without Logs & 1,243,655 (5.4\%) & 188.4 (3.7\%) & 90,422,493 (73.8\%) & 114.7 (12.8\%) & 770,885,864 (62.7\%) & 2,002.8 (28.2\%) \\
\midrule
\textbf{Total} & \textbf{22,830,229 (100.0\%)} & \textbf{5,141.6 (100.0\%)} & \textbf{122,454,486 (100.0\%)} & \textbf{894.7 (100.0\%)} & \textbf{1,228,926,146 (100.0\%)} & \textbf{7,110.45 (100.0\%)} \\ 
\bottomrule
\end{tabular}
\end{adjustbox}
\end{center}
\caption{Overview of number of transactions and block producer earnings in ETH for each identified high-level state-invariant transaction category per blockchain.}
\label{tab:spam-txs}
\end{table*}

\begin{figure}[t]
    \centering
    \begin{subfigure}[b]{0.49\textwidth}
        \centering
        \includegraphics[width=\textwidth]{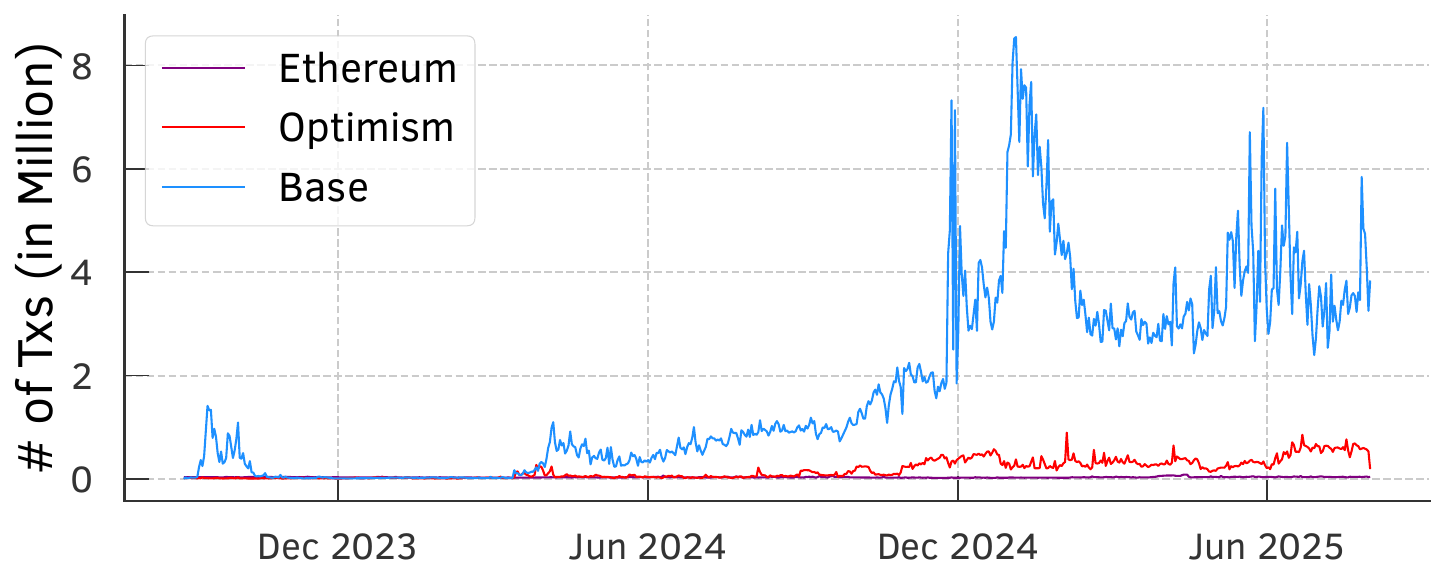}
        \caption{Number of State-Invariant Transactions}
        \label{fig:panel-a}
    \end{subfigure}
    \begin{subfigure}[b]{0.49\textwidth}
        \centering
        \includegraphics[width=\textwidth]{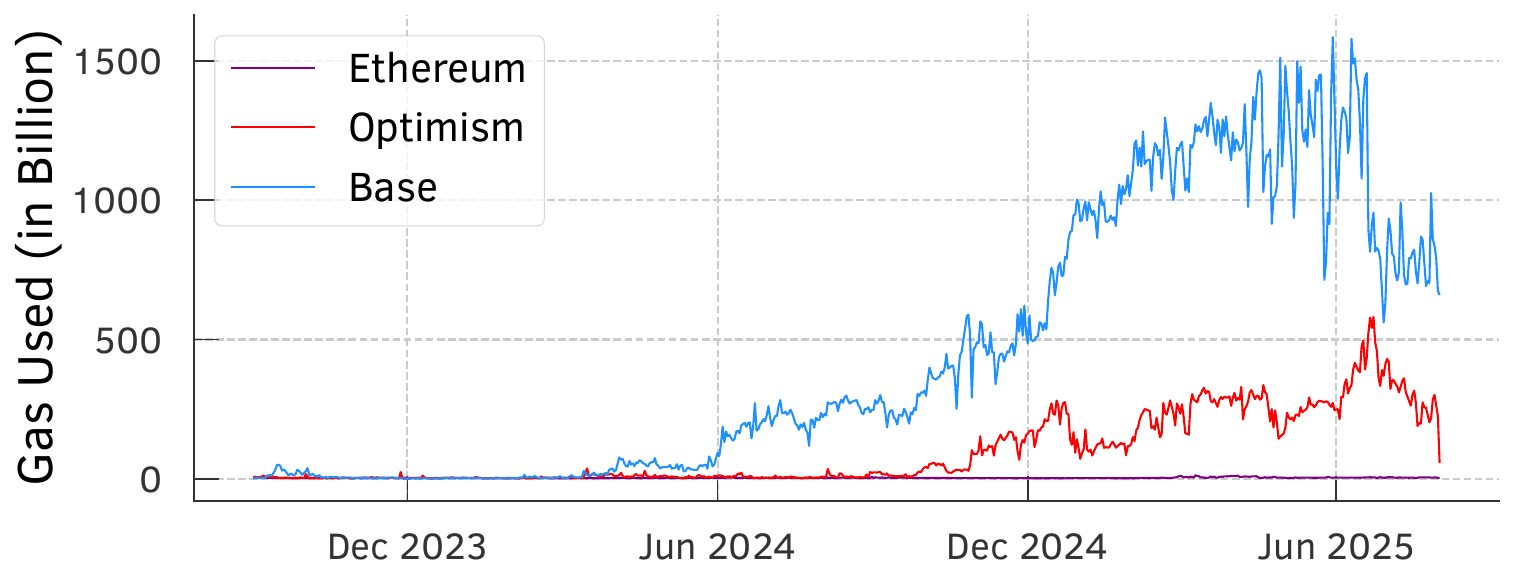}
        \caption{Gas Used}
        \label{fig:panel-b}
    \end{subfigure}
    \begin{subfigure}[b]{0.49\textwidth}
        \centering
        \includegraphics[width=\textwidth]{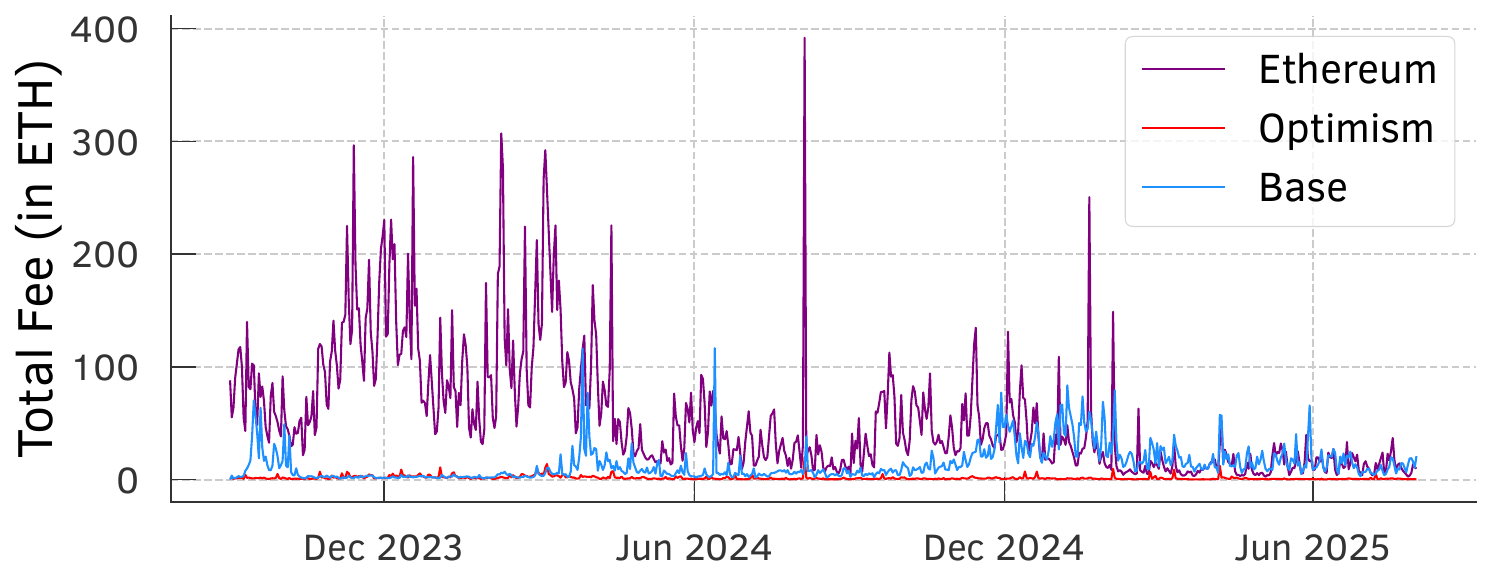}
        \caption{Total Fee (with Base Fee)}
        \label{fig:panel-c}
    \end{subfigure}
    \begin{subfigure}[b]{0.49\textwidth}
        \centering
        \includegraphics[width=\textwidth]{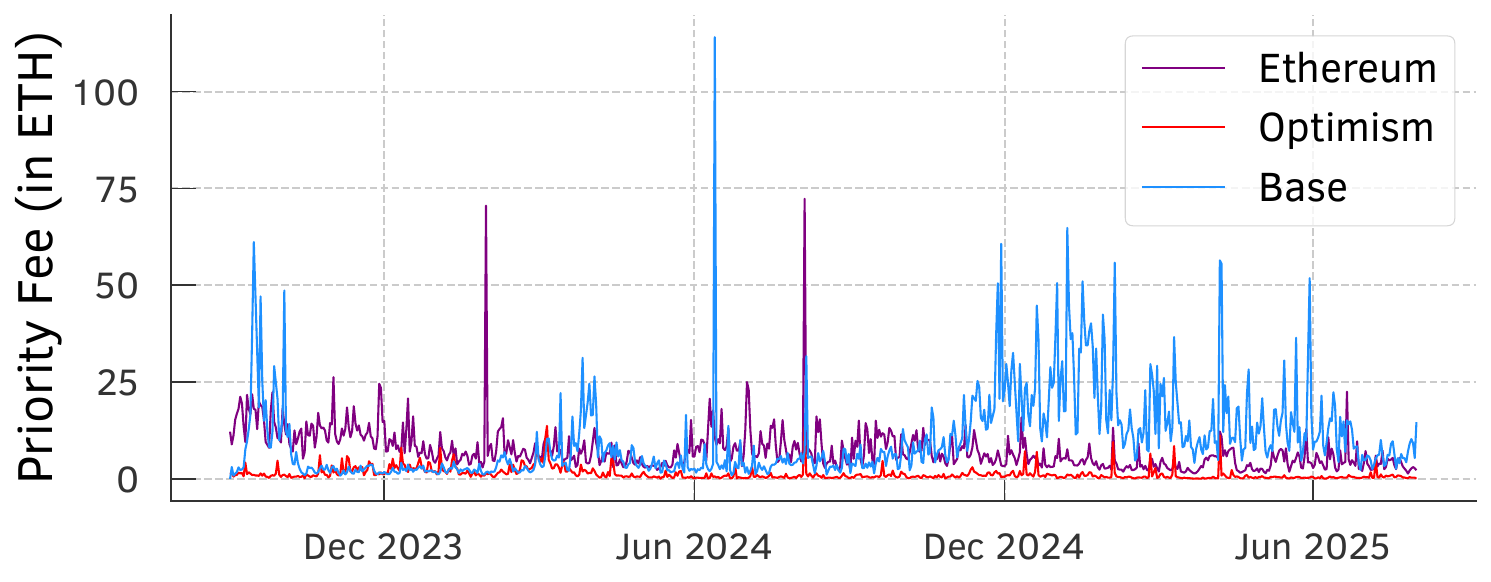}
        \caption{Priority Fee (without Base Fee)}
        \label{fig:panel-d}
    \end{subfigure}

    \caption{Overview of state-invariant transactions over time across Ethereum, Optimism, and Base in terms of number of transactions, gas used, total transaction fees, and priority fees.}
    \label{fig:spam-txs-statistics}
\end{figure}

Figure~\ref{fig:spam-txs-statistics} shows the daily number of state-invariant transactions on Ethereum, Optimism, and Base over our study period.
We observe that both the volume of state-invariant transactions and the associated gas consumption are substantially higher on Optimism and Base than on Ethereum. 
Despite this increased activity, the priority fees paid by spam transactions on the Layer-2 networks remain relatively low, reflecting their significantly lower transaction costs.

\begin{figure}[t]
    \centering
    \begin{subfigure}[b]{0.49\textwidth}
        \centering
        \includegraphics[width=\textwidth]{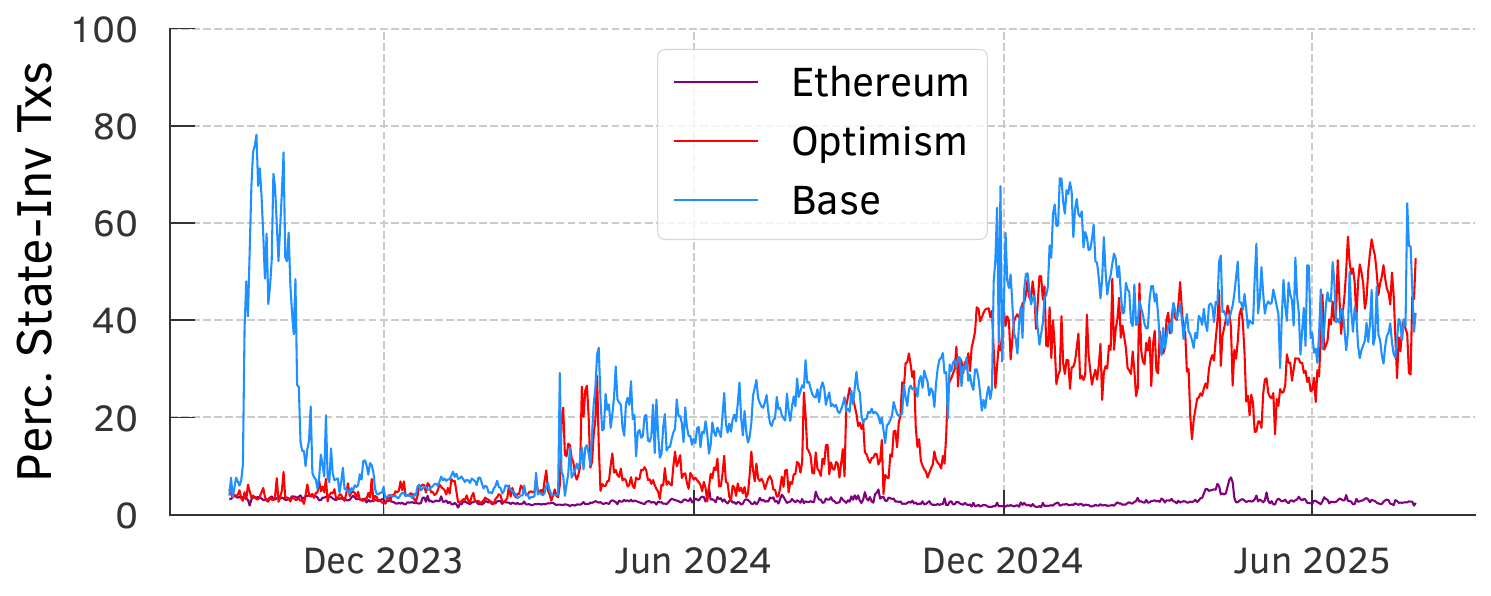}
        \caption{Proportion of Daily Transactions}
        \label{fig:panel-a}
    \end{subfigure}
    \begin{subfigure}[b]{0.49\textwidth}
        \centering
        \includegraphics[width=\textwidth]{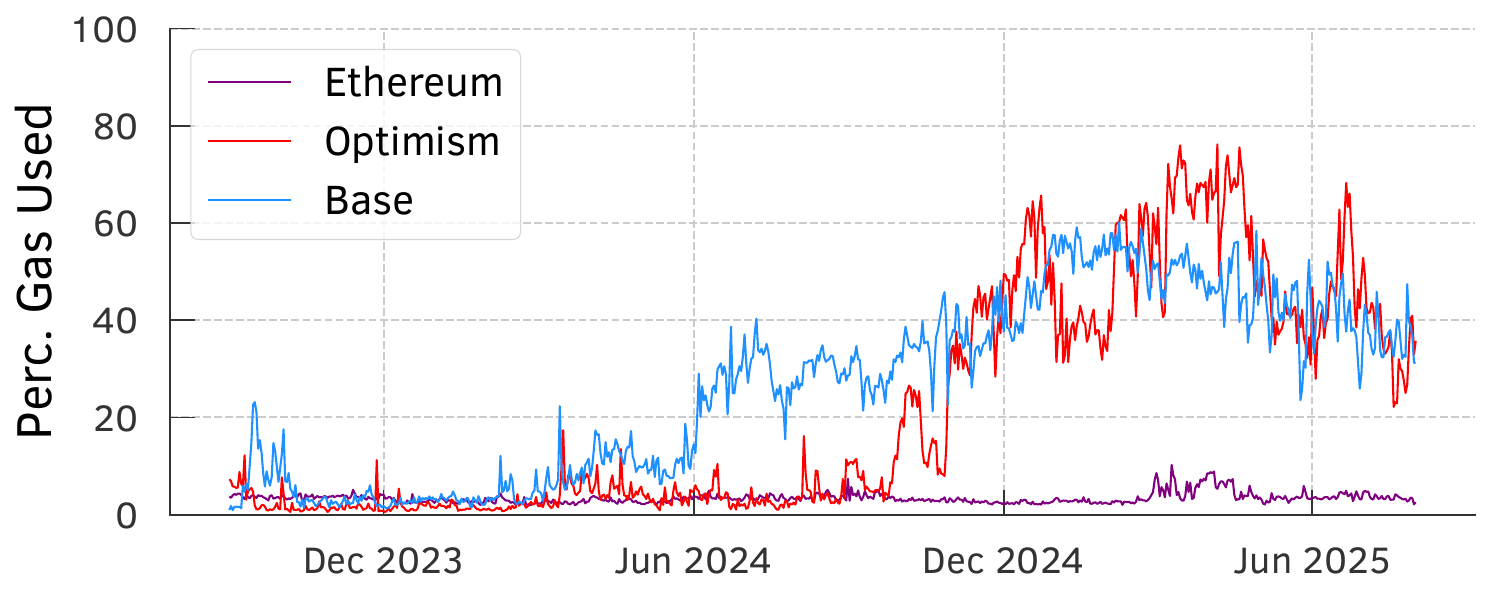}
        \caption{Proportion of Daily Gas Used}
        \label{fig:panel-b}
    \end{subfigure}
    \caption{Overview of daily proportion of state-invariant transactions and gas consumption relative to the total transaction volume and gas usage.}
    \label{fig:perc-spam-txs}
\end{figure}

Figure~\ref{fig:perc-spam-txs} presents the fraction of state-invariant transactions relative to all transactions on each network. We observe that the share of state-invariant transactions on Optimism and Base increases markedly following the Dencun upgrade in March 2024. During much of the post-Dencun period, state-invariant transactions consistently account for approximately 40--50\% of the daily number of transactions and gas consumed on both Layer-2 blockchains. In contrast, Ethereum consistently exhibits a much smaller proportion of state-invariant transactions throughout the study period.

\subsubsection{Failed Transactions}

A closer examination of the identified state-invariant transactions reveals two broad categories: \emph{failed transactions} (\texttt{tx.status = 0}) and \emph{successful transactions} (\texttt{tx.status = 1}).
We partition failed transactions into two classes: \emph{Out of Gas} failures and \emph{reverted} transactions.
A transaction is classified as out-of-gas if the gas consumed equals the transaction's gas limit, indicating that execution exhausted the available gas.
All remaining failed transactions are classified as reverted, meaning that execution terminated for reasons other than running out of gas.

The composition of state-invariant transactions differs substantially between Ethereum and the Layer-2 networks. On Ethereum, approximately 69\% of state-invariant transactions are failed transactions, compared to only 19\% on Optimism and 32\% on Base.
Despite accounting for a minority of state-invariant transactions on the Layer-2s, failed transactions are responsible for a disproportionately large share of the fees paid. 
Specifically, failed transactions account for 88.3\%, 80.9\%, and 67.7\% of the priority fees (i.e., the total transaction fee excluding the base fee) paid by state-invariant transactions on Ethereum, Optimism, and Base, respectively.
This indicates that failed transactions are, on average, paying more than successful state-invariant transactions, particularly on Optimism and Base, where they consume a much larger share of fees than their share of transaction volume would suggest.

\subsubsection{Successful Transactions}

Although all failed transactions are classified as spam under our methodology, only a subset of successful transactions satisfy our definition of spam. 
The proportion of successful state-invariant transactions varies substantially across the three networks. On Ethereum, only 30.9\% of state-invariant transactions are successful, whereas this proportion increases to 80.6\% on Optimism and 67.4\% on Base.
Despite their prevalence on the Layer-2s, successful state-invariant transactions account for a relatively small fraction of the fees paid by all state-invariant transactions. Specifically, they contribute only 11.7\%, 19.1\%, and 32.3\% of the total priority fees on Ethereum, Optimism, and Base, respectively. This suggests that successful state-invariant transactions are, on average, pay less than failed state-invariant transactions.

We further classify successful state-invariant transactions based on whether they emit event logs during execution. 
Here, we again observe a clear distinction between Ethereum and the Layer-2 networks. On Ethereum, approximately 83\% of successful state-invariant transactions emit at least one event log.
In contrast, only 8\% and 5\% of successful state-invariant transactions emit logs on Optimism and Base, respectively. 
Thus, the overwhelming majority of successful state-invariant transactions on the Layer-2s complete without producing any event logs.

Examining the Ethereum state-invariant transactions that emit event logs reveals that the majority generate at least one \texttt{Transfer} event. Among these, approximately 10\% correspond to \emph{self-transfers}, where the sender and recipient are the same address. In every transaction in this category, the \texttt{to} field references an ERC-20 token contract directly, indicating that the transaction invokes the token contract without transferring tokens between distinct accounts.
We also identify a large number of transactions that repeatedly issue the same ERC-20 approval even though the requested allowance has already been granted. These redundant approval transactions account for more than 37\% of all log-emitting state-invariant transactions on Ethereum. As with self-transfers, the vast majority of these transactions invoke an ERC-20 token contract directly through the \texttt{to} field.
We observe similar transactions and pattern in Optimism and Base as well; however, not to the same extent proportionally.
This behavior points to inefficiencies in wallet software or decentralized applications, which repeatedly submit approval transactions despite no change in the token allowance. As a result, users incur transaction fees for operations that have no effect on the underlying on-chain state.

\subsubsection{Top Contracts}

Examining the \texttt{to} field of state-invariant transactions on Ethereum reveals that six of the ten most frequently invoked contracts are DEX routers, while three are ERC-20 token contracts. These ten contracts account for 31\% of all state-invariant transactions on Ethereum. This suggests that a substantial fraction of state-invariant activity is driven by trading-related operations, such as swaps that ultimately revert due to slippage or other execution constraints. The prominence of ERC-20 token contracts indicates that many state-invariant transactions correspond to operations such as zero-value transfers, self-transfers, or redundant approvals.

The picture is markedly different on the Layer-2 networks. On Optimism, the ten most frequently called contract addresses account for approximately 40\% of all state-invariant transactions, and all are associated with MEV strategies such as arbitrage and liquidations. Remarkably, 99.9\% of these transactions are successful, yet they emit no event logs, indicating that execution completes without producing any observable state changes or events. On Base, six of the ten most frequently invoked contracts are associated with MEV bots, accounting for approximately 13\% of all identified state-invariant transactions. These results reinforce our earlier finding that state-invariant transactions on Layer-2 networks are dominated by MEV-related activity, whereas Ethereum exhibits a more diverse mix of DEX trading and token transfers.

\subsection{Speculative MEV}

Following the heuristics introduced in Section~\ref{sec:methodology}, we classify successful arbitrage and liquidation transactions as speculative or non-speculative. Based on these labels, we then categorize MEV bots as speculative, non-speculative, or hybrid. Finally, we associate identified state-invariant transactions with one of these categories by matching their destination addresses against our dataset of classified MEV bots.

\subsubsection{Overview}

\begin{table*}
\begin{center}
\begin{adjustbox}{max width=\linewidth}
\begin{tabular}{l r r r r r r r r r}
\toprule
 & \multicolumn{3}{c}{\textbf{Ethereum}} & \multicolumn{3}{c}{\textbf{Optimism}} & \multicolumn{3}{c}{\textbf{Base}} \\
 \cline{2-10} \noalign{\vskip 3pt}
 \textbf{MEV Bot Strategy} & \multicolumn{1}{c}{\textbf{Bots}} & \multicolumn{1}{c}{\textbf{State-Inv. Txs}} & \textbf{Suc. Txs} & \multicolumn{1}{c}{\textbf{Bots}} & \multicolumn{1}{c}{\textbf{State-Inv. Txs}} & \textbf{Suc. Txs} & \multicolumn{1}{c}{\textbf{ Bots}} & \multicolumn{1}{c}{\textbf{State-Inv. Txs}} & \textbf{Suc. Txs} \\
\midrule

\hspace{0.0cm} \textbf{Speculative} & 582 & 41,968 & 292,415 & 2,565 & 83,760,338 & 1,733,462 & 9,370 & 681,985,254 & 8,535,187 \\
\hspace{0.2cm} Arbitrage & 498 & 28,154 & 291,848  & 2,132 & 81,564,065 & 1,727,972 & 8,813 & 658,483,710 & 8,525,443 \\
\hspace{0.2cm} Liquidation & 84 & 13,814 & 567  & 433 & 2,196,273 & 5,490 & 557 & 23,501,544 & 9,744 \\
\hspace{0.0cm} \textbf{Non-Speculative} & 836 & 414,821 & 305,427 & 159 & 223,969 & 46,545 & 728 & 1,348,933 & 376,847 \\
\hspace{0.2cm} Arbitrage & 794 & 414,429 & 305,173 & 142 & 223,017 & 46,400 & 693 & 1,327,644 & 376,405 \\
\hspace{0.2cm} Liquidation & 54 & 435 & 254 & 17 & 952 & 145 & 35 & 21,289 & 442 \\
\hspace{0.0cm} \textbf{Hybrid} & 468 & 227,892 & 1,093,707 & 278 & 5,071,709 & 2,569,587 & 673 & 32,930,392 & 555,6975 \\
\hspace{0.2cm} Arbitrage & 388 & 223,701 & 1,064,144 &  221 & 4,792,528 & 2,558,803 & 581 & 32,406,533 & 5,529,470 \\
\hspace{0.2cm} Liquidation & 118 & 29,795 & 29,563 & 61 & 1,315,749 & 10,784 & 96 & 5,652,728 & 27,505 \\
\midrule
\textbf{Total} & \textbf{1,845} & \textbf{667,139} & \textbf{1,691,549} & \textbf{2,999} & \textbf{89,052,271} & \textbf{4,349,594} & \textbf{10,769} & \textbf{716,241,462} & \textbf{14,469,009} \\ 
\bottomrule
\end{tabular}
\end{adjustbox}
\end{center}
\caption{Overview of state-invariant transactions across different MEV strategies. Speculative MEV dominates MEV-related state-invariant transactions on Optimism and Base, whereas non-speculative MEV is predominant on Ethereum. Overall, MEV-related state-invariant transactions account for approximately 73\% of all state-invariant transactions on Optimism but only about 3\% on Ethereum.
}
\label{label:speculative-non-speculative-hybrid-table}
\end{table*}

Overall, we find that only 3\% of state-invariant transactions on Ethereum can be attributed to the identified MEV bots. In contrast, this proportion is substantially higher on the Layer-2s, with 73\% of state-invariant transactions on Optimism and 60\% on Base corresponding to identified bots.
Table~\ref{label:speculative-non-speculative-hybrid-table} summarizes our findings for speculative, non-speculative, and hybrid bots, following the definitions and methodology introduced in the previous section.

On Ethereum, we identify 1,845 MEV bots that execute approximately 1.6 million successful transactions and 0.67 million state-invariant transactions, corresponding to an overall success rate of 72\%. Most state-invariant transactions associated with bots (approximately 62\%) originate from non-speculative arbitrage bots. Notably, 87\% of these state-invariant transactions are generated by just two bots, indicating a high concentration of unsuccessful activity.

The picture is markedly different on the Layer-2s, where success rates are substantially lower. On Optimism, we identify approximately 4.3 million successful MEV transactions compared to nearly 89 million state-invariant transactions, yielding a success rate of only 4.8\%. Unlike Ethereum, the overwhelming majority of state-invariant transactions (91\%) are generated by speculative arbitrage bots. The five largest speculative arbitrage bots account for nearly 45\% of all state-invariant arbitrage transactions, despite contributing only 14\% of successful arbitrage transactions.

We observe a similar pattern on Base. We identify approximately 14.4 million successful MEV transactions alongside 716 million state-invariant transactions, corresponding to a success rate of just 2\%. As on Optimism, speculative arbitrage bots dominate unsuccessful activity, accounting for 92\% of all state-invariant transactions attributed to MEV bots. The five largest speculative arbitrage bots are responsible for nearly 15\% of all state-invariant arbitrage transactions on Base, while contributing only 1.5\% of successful arbitrage transactions.

Although we observe speculative liquidations in the wild, they are considerably less common than speculative arbitrages, which is expected given the relative scarcity of liquidation opportunities.

\subsubsection{Private Transactions} 

Prior work~\cite{Weintraub-IMC22} has extensively studied the use of private transaction submission by MEV bots on Ethereum through mechanisms such as private mempools and direct submission to block proposers. In this work, we examine whether state-invariant transactions from MEV bots were broadcasted through the public mempool.
To do so, we leverage public data from the Flashbots Mempool Dumpster project \cite{dumpster}, which records transactions observed in the public mempool, including those that were never included on-chain. During our study period, we identify 24,218,195 public mempool transactions involving the Ethereum MEV bots in our dataset. However, only 4,164,112 of these transactions were eventually included in a block, implying that just 17.2\% of publicly broadcast transactions are successfully included on-chain.

Approximately 70\% of these 4.16 million publicly submitted and subsequently included transactions originate from a single MEV bot: \texttt{0xf3de3c0d654fda23dad170f0f320a92172509127}. Among these 4.16 million transactions, 577,119 are state-invariant, representing approximately 87\% of all state-invariant transactions attributed to the identified MEV bots. Furthermore, two bots alone account for 54\% of these publicly submitted state-invariant transactions.
Furthermore, out of the almost 1.7 million identified successful arbitrage and liquidation transactions, only 20,724 were sent publicly accounting for only around 1.2\%. 
Hence, a substantial fraction of the MEV-related state-invariant transactions observed on Ethereum can be attributed to bots that submit transactions through the public mempool rather than via private bundle submission services, thereby forgoing revert protection. In contrast, only a relatively small fraction of successful MEV transactions are submitted through the public mempool.

These findings highlight the important role of Ethereum's existing MEV transaction supply chain ecosystem in reducing on-chain spam. 
A large fraction of publicly broadcast transactions never reach the blockchain, suggesting that they are filtered out through competition before block inclusion. 
In contrast, no comparable off-chain auction mechanism exists on Optimism or Base. 
Rollups have a private mempool and transactions submitted to the sequencer are generally included on-chain unless they are explicitly canceled. 
As a result, unsuccessful transaction attempts are far more likely to appear on-chain, contributing to the substantially higher volume of state-invariant transactions observed on these networks.

\begin{figure}[t]
    \centering
    \begin{subfigure}[b]{0.325\textwidth}
        \centering
        \includegraphics[width=\textwidth]{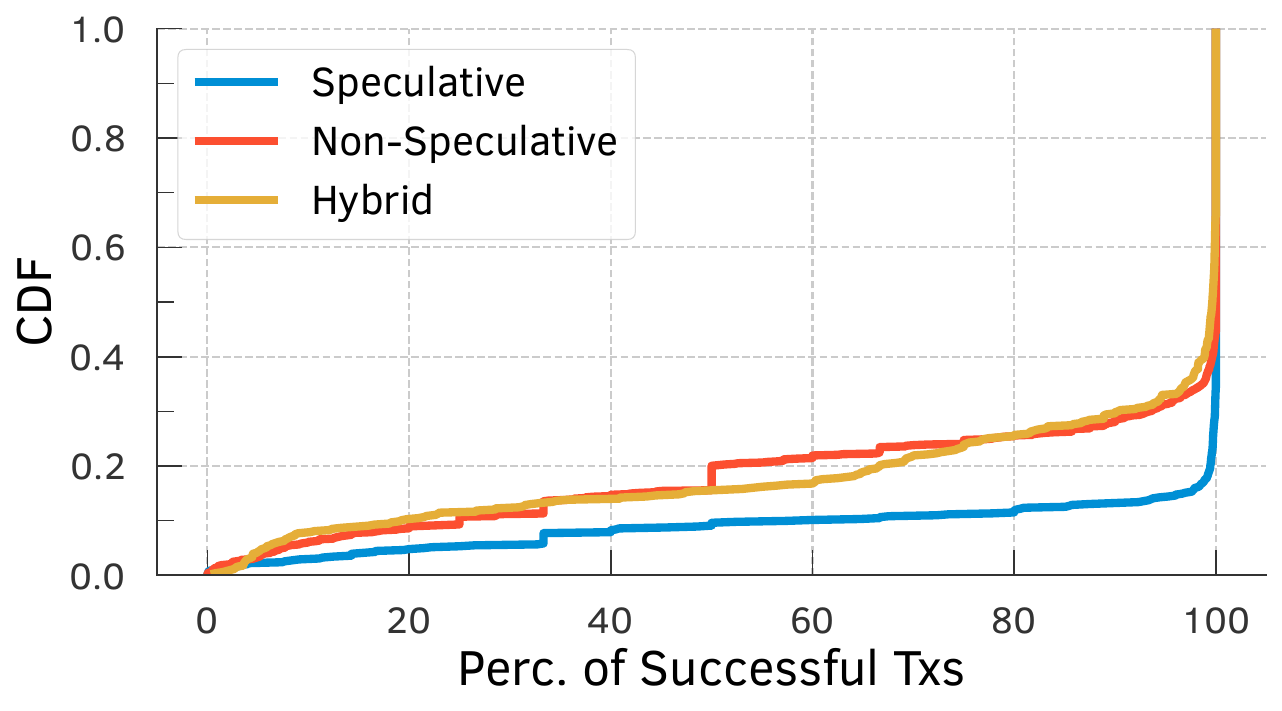}
        \caption{Ethereum}
        \label{fig:eth_perc_suc}
    \end{subfigure}
    \begin{subfigure}[b]{0.325\textwidth}
        \centering
        \includegraphics[width=\textwidth]{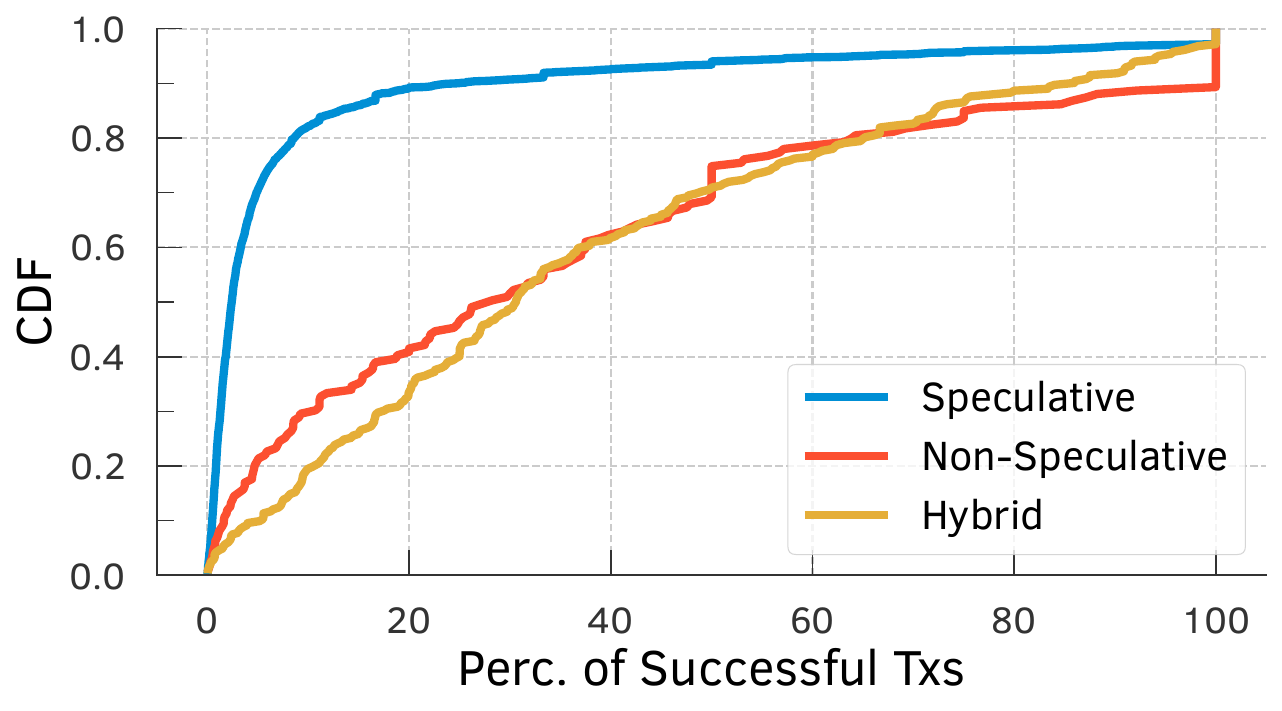}
        \caption{Optimism}
        \label{fig:opt_perc_suc}
    \end{subfigure}
    \begin{subfigure}[b]{0.325\textwidth}
        \centering
        \includegraphics[width=\textwidth]{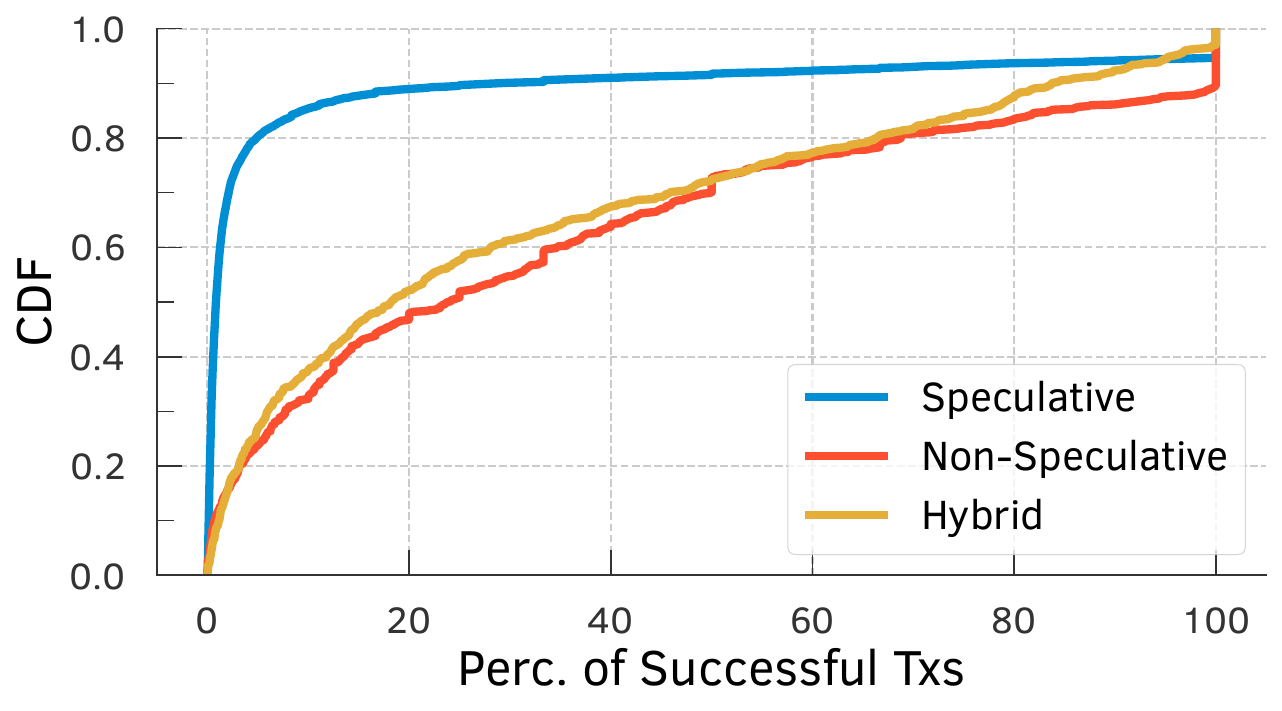}
        \caption{Base}
        \label{fig:base_perc_suc}
    \end{subfigure}
    \caption{Percentage of successful MEV transactions per MEV bot strategy across the three blockchains.}
    \label{fig:speculative_perc_success}
\end{figure}

\subsubsection{Success Rate}

We next analyze the success rates of speculative, non-speculative, and hybrid MEV bots across the three blockchains in Figure~\ref{fig:speculative_perc_success}. To compute each bot's success rate, we consider both the number of successful MEV transactions it executes and the number of state-invariant transactions it submits. The success rate is defined as the fraction of successful MEV transactions out of the total number of successful MEV and state-invariant transactions submitted by the bot.

On Ethereum, speculative arbitrage bots exhibit the highest success rates. More than 80\% of these bots successfully execute nearly all of their submitted transactions, compared to roughly 60\% for both non-speculative and hybrid bots. This result is consistent with Ethereum's mature private transaction ecosystem that offers revert protection and atomicity to searchers. As discussed previously, speculative arbitrage transactions are frequently submitted through private channels, allowing bots to avoid unsuccessful transactions being exposed on-chain. Consequently, non-speculative bots account for the majority of state-invariant transactions observed on Ethereum.
In contrast, speculative arbitrage bots on Optimism and Base exhibit much lower success rates. On both networks, over 90\% of speculative bots successfully execute fewer than 10\% of their transactions, indicating that the vast majority of their attempts result in state-invariant transactions. Interestingly, despite this poor performance among speculative bots, approximately 30\% of non-speculative and hybrid bots on both Optimism and Base achieve success rates exceeding 50\%. Nevertheless, even these bots generate a substantial number of state-invariant transactions.

\subsubsection{Calldata}

To better understand the execution strategies employed by MEV bots, we analyze the calldata of their transactions.
Table~\ref{tab:calldata} shows that speculative arbitrage bots use substantially larger calldata than other categories of MEV bots. 
This suggests that these bots encode a significant portion of their execution logic or trading strategy directly within the transaction calldata, which is then interpreted by a generic on-chain contract.
We also examine the extent to which bots reuse identical calldata across transactions. 
We quantify calldata reuse by first computing the calldata diversity $d$ of each bot as the ratio of unique calldata values to the total number of calldata values. The reuse ratio is then obtained as $1 - d$. Finally, we average the reuse ratios across all bots of the same MEV strategy on each blockchain.

We find that speculative arbitrage bots consistently exhibit the highest calldata reuse rates across all three blockchains. The effect is particularly pronounced on the Layer-2s, where speculative arbitrage bots reuse calldata on average in 82\% of transactions on Optimism and 54\% on Base. This high degree of reuse indicates that these bots frequently execute the same underlying strategy while varying only a small number of transaction parameters or none at all.
Speculative liquidation bots also exhibit higher calldata reuse than their non-speculative counterparts. 
However, their reuse rates remain considerably lower than those of speculative arbitrage bots.
A likely explanation is that liquidation transactions are inherently more complex, typically involving multiple sequential operations, including debt repayment, collateral acquisition, and subsequent asset swaps, which require transaction-specific calldata and therefore reduce opportunities for reuse.

\begin{table}[]
    \centering
    \begin{adjustbox}{max width=\columnwidth}
    \begin{tabular}{l l r r r r r r r r r r}
    \toprule
    & & \multicolumn{5}{c}{\textbf{Arbitrage}} & \multicolumn{5}{c}{\textbf{Liquidation}} \\
    \cline{3-12} \noalign{\vskip 3pt}
       \textbf{Chain} & \textbf{MEV Bot Type} & \multicolumn{1}{c}{\textbf{Min}} & \multicolumn{1}{c}{\textbf{Mean}} & \multicolumn{1}{c}{\textbf{Median}} & \multicolumn{1}{c}{\textbf{Max}} & \multicolumn{1}{c}{\textbf{Reuse}} & \multicolumn{1}{c}{\textbf{Min}} & \multicolumn{1}{c}{\textbf{Mean}} & \multicolumn{1}{c}{\textbf{Median}} & \multicolumn{1}{c}{\textbf{Max}} & \multicolumn{1}{c}{\textbf{Reuse}} \\ 
    \midrule
    \multirow{3}{*}{\text{Ethereum}} & Speculative & 4 & 816 & 220 & 38,020 & 3.92\% & 4 & 479 & 228 & 9,503 & 1.79\% \\
        & Non-Speculative & 13 & 332 & 155 & 17,092 & 0.95\% & 100 & 819 & 455 & 5,732 & 0.00\% \\
        & Hybrid & 13 & 327 & 128 & 33,490 & 0.83\% & 108 & 989 & 612 & 35,565 & 0.90\% \\
    \midrule
    \multirow{3}{*}{\text{Base}} & Speculative & 0 & 411 & 90 & 62,052 & 54.87\% & 0 & 390 & 107 & 12,612 & 7.16\% \\
        & Non-Speculative & 19 & 275 & 146 & 8,356 & 2.88\% & 100 & 656 & 612 & 6,683 & 0.00\% \\
        & Hybrid & 4 & 273 & 147 & 35,857 & 4.50\% & 41 & 590 & 410 & 7,876 & 0.01\% \\
    \midrule
    \multirow{3}{*}{\text{Optimism}} & Speculative &  0 & 341 & 90 & 58,244 & 82.35\% & 0 & 222 & 42 & 6,564 & 6.75\% \\
        & Non-Speculative & 13 & 215 & 123 & 6,724 & 1.84\% & 100 & 622 & 619 & 3,284 & 0.00\% \\
        & Hybrid & 4 & 193 & 160 & 7,396 & 3.55\% & 0 & 589 & 560 & 9,828 & 3.80\% \\
    \bottomrule    
    \end{tabular}
    \end{adjustbox}
    \caption{Statistical summary of calldata lengths measured in number of bytes and average calldata reuse ratio for MEV bots across multiple bot types and blockchains.}
    \label{tab:calldata}
\end{table}

\subsubsection{Profitability}

We next examine the profitability of MEV bots by comparing the revenue generated from arbitrage and liquidation opportunities with the transaction fees they incur. Table~\ref{label:speculative-non-speculative-hybrid-profits} summarizes the results.
When considering only successful MEV transactions, we find that 75\% of Ethereum MEV bots are profitable. These profitable bots are responsible for 97\% of all successful MEV transactions, while accounting for only 43\% of the state-invariant transactions attributed to MEV bots. On the Layer-2 networks, 60\% of MEV bots on Optimism and 82\% on Base are profitable under the same metric. As on Ethereum, profitable bots dominate successful MEV activity, accounting for 97\% and 98\% of successful MEV transactions on Optimism and Base, respectively.

This analysis considers only the transaction fees associated with successful MEV transactions. State-invariant transactions also incur transaction fees, particularly on Layer-2 blockchains where nearly every submitted transaction is ultimately included on-chain. 
Hence, we extend our analysis by incorporating the fees paid for state-invariant transactions into each bot's total operating costs.

On Ethereum, accounting for these additional costs has only a modest impact: the proportion of profitable bots decreases from 75\% to 73\%. This is expected, as relatively few state-invariant transactions are included on-chain. In contrast, the effect is dramatic on the Layer-2s. On Optimism, the proportion of profitable bots falls from 60\% to just 24\%. The largest impact is observed among speculative bots, where the number of profitable bots declines from 1,489 to 547, a reduction of more than 63\%. Similarly, on Base, the proportion of profitable bots decreases from 82\% to 28\%, with speculative arbitrage bots again experiencing the largest decline, falling from 7,774 to 2,454 profitable bots (a 68\% reduction).
These findings are consistent with our earlier observations that speculative bots generate the overwhelming majority of state-invariant transactions on the Layer-2s—more than 93\% on Optimism and 95\% on Base. Consequently, the fees incurred by unsuccessful transaction attempts constitute a substantial component of their operating costs, significantly reducing their overall profitability.

\begin{table*}
\begin{center}
\begin{adjustbox}{max width=\linewidth}
\begin{tabular}{l r r r r r r r r r}
 & \multicolumn{3}{c}{\textbf{Ethereum}} & \multicolumn{3}{c}{\textbf{Optimism}} & \multicolumn{3}{c}{\textbf{Base}} \\
 \cline{2-10} \noalign{\vskip 3pt}
 \textbf{MEV Bot Strategy} & \multicolumn{1}{r}{\textbf{Prof. Bots}} & \multicolumn{1}{c}{\textbf{Bots after Cost}} & \textbf{Profits [ETH]} & \multicolumn{1}{r}{\textbf{Prof. Bots}} & \multicolumn{1}{c}{\textbf{Bots after Cost}} & \textbf{Profits [ETH]} & \multicolumn{1}{r}{\textbf{Prof. Bots}} & \multicolumn{1}{c}{\textbf{Bots after Cost}} & \textbf{Profits [ETH]} \\
\midrule

\hspace{0.0cm} \textbf{Speculative} & \textbf{441 (75.8\%)} & \textbf{429 (73.7\%)} & \textbf{5,822.20} & \textbf{1,489 (58.1\%)} & \textbf{547 (21.3\%)} & \textbf{75.03} & \textbf{7,774 (83.0\%)} & \textbf{2,454 (26.2\%)} & \textbf{361.93} \\
\hspace{0.2cm} Arbitrage & 419 (84.1\%) & 409 (82.1\%) & 5,169.41 & 1,205 (56.5\%) & 280 (13.1\%) & -31.26 & 7,359 (83.5\%) & 2,058 (23.4\%) & -28.07 \\
\hspace{0.2cm} Liquidation & 22 (26.2\%) & 20 (23.8\%) & 652.79 & 284 (65.6\%) & 267 (61.7\%) & 106.29 & 415 (74.5\%) & 396 (71.1\%) & 390.00 \\

\hspace{0.0cm} \textbf{Non-Speculative} & \textbf{563 (66.4\%)} & \textbf{528 (62.3\%)} & \textbf{35,304.05} & \textbf{87 (54.7\%)} & \textbf{41 (25.8\%)} & \textbf{-3.12} & \textbf{486 (66.8\%)} & \textbf{198 (27.2\%)} & \textbf{231.30} \\
\hspace{0.2cm} Arbitrage & 531 (66.9\%) & 500 (63.0\%) & 34,961.52 & 77 (54.2\%) & 31 (21.8\%) & -12.45 & 457 (65.9\%) & 174 (25.1\%) & 73.53 \\
\hspace{0.2cm} Liquidation & 32 (59.3\%) & 28 (51.9\%) & 342.53 & 10 (58.8\%) & 10 (58.8\%) & 9.33 & 29 (82.9\%) & 24 (68.6\%) & 157.77 \\

\hspace{0.0cm} \textbf{Hybrid} & \textbf{444 (87.9\%)} & \textbf{430 (85.1\%)} & \textbf{248,456.56} & \textbf{221 (78.4\%)} & \textbf{137 (48.6\%)} & \textbf{505.26} & \textbf{576 (85.1\%)} & \textbf{343 (50.7\%)} & \textbf{2,215.05} \\
\hspace{0.2cm} Arbitrage & 340 (87.9\%) & 330 (85.3\%) & 221,218.58 & 166 (75.1\%) & 89 (40.3\%) & 107.31 & 489 (84.2\%) & 267 (46.0\%) & 1,437.04 \\
\hspace{0.2cm} Liquidation & 104 (88.1\%) & 100 (84.7\%) & 27,237.98 & 55 (90.2\%) & 48 (78.7\%) & 397.95 & 87 (90.6\%) & 76 (79.2\%) & 778.01 \\
\midrule

\textbf{Total} & \textbf{1,448 (75.1\%)} & \textbf{1,387 (71.9\%)} & \textbf{289,582.81} & \textbf{1,797 (59.8\%)} & \textbf{725 (24.1\%)} & \textbf{577.17} & \textbf{8,836 (82.1\%)} & \textbf{2,995 (27.8\%)} & \textbf{2,808.28} \\ 
\bottomrule
\end{tabular}
\end{adjustbox}
\end{center}
\caption{Overview of the profitability of speculative, non-speculative, and hybrid MEV bots before and after accounting for the transaction fees incurred by state-invariant transactions. Profitability is first computed using arbitrage and liquidation revenue minus the fees paid by successful MEV transactions only, and then recomputed after including the fees paid by state-invariant transactions. While the additional costs have only a marginal effect on profitability on Ethereum, they substantially reduce profitability on Optimism and Base, with speculative MEV bots becoming unprofitable in aggregate.
}
\label{label:speculative-non-speculative-hybrid-profits}
\end{table*}

\subsection{Non-MEV State-Invariant Transactions}

Previous studies identify ``spam'' transactions by first detecting arbitrage bots using heuristic-based techniques and subsequently classifying the transactions generated by these bots as spam. Consequently, their analyses are restricted to transactions originating from bots that satisfy the chosen heuristics, inherently excluding spam transactions that do not meet these identification criteria. While such approaches have provided valuable insights into the impact of arbitrage bots on blockchain ecosystems, they do not capture the full spectrum of spam activity.
In contrast, our methodology identifies state-invariant transactions independently of the entities that generate them. Hence, it captures a substantially broader set of spam transactions, extending beyond those produced by arbitrage bots.
This discrepancy cannot be explained solely by failed transactions. Even after excluding state-invariant transactions that fail, our methodology continues to identify a substantial number of spam transactions missed by existing approaches. Specifically, we find that 27\% of state-invariant transactions on Optimism and 41\% on Base cannot be attributed to arbitrage or liquidation bots identified by prior methodologies. On Ethereum, only 3\% of state-invariant transactions are associated with arbitrage or liquidation activity, indicating that existing heuristic-based approaches miss a substantial fraction of spam transactions.

As discussed before, on Ethereum, the majority of state-invariant transactions are failed transactions. Among the ten most frequently targeted non-MEV addresses by failed transactions, seven are decentralized exchange (DEX) routers and three are ERC-20 token contracts. Although transactions interacting with these contracts rarely fail under normal operation, the corresponding state-invariant transactions almost always revert.

To identify non-MEV bots on Ethereum, we consider destination contracts that (i) have generated at least 10,000 state-invariant transactions and (ii) do not have verified source code on Etherscan. Using this heuristic, we identify 71 bot contracts. Among the ten most active, seven are token-swapping bots. We identify these by randomly sampling 100 successful transactions per contract and classifying a contract as a swap bot if more than 50\% of the sampled transactions emit swap events. The remaining three contracts are labeled by Etherscan as phishing-related.

Applying the same methodology to Optimism, we find that nine most active non-MEV bot contracts are swap bots, while one is label as being associated with phishing activity by Optimistic Etherscan.
Base exhibits a markedly different pattern. Only two of the ten most active non-MEV bot contracts are classified as swap bots. The remaining eight contracts do not emit any events, yet they modify contract storage without performing any externally observable meaningful action. For each of these contracts, more than 99\% of their transactions are classified as state-invariant, with only a small fraction producing any state changes. Collectively, these eight contracts account for approximately 72 million or 6\% of all state-invariant transactions, demonstrating that non-MEV bots constitute a major source of state-invariant activity on Base.

\subsection{Address Poisoning}

Having characterized non-MEV bots, we next examine a distinct class of transactions identified by our state-invariant transaction methodology. We find that 31\% of state-invariant transactions emitting \texttt{Transfer} events on Ethereum are `zero-transfer' transactions, in which the transferred amount is zero. Although these transactions emit standard ERC-20 \texttt{Transfer} events, they do not result in any meaningful token balance changes. Upon further inspection, we observe that eight of the ten most frequently targeted destination addresses are labeled as `Phishing' by Etherscan. This behavior is consistent with the address poisoning attacks described by Tsuchiya \ea~\cite{tsuchiya2025blockchainaddresspoisoning}.

Our methodology identifies 17,240,713 state-invariant transactions, of which 1,760,004 are contained in the published address poisoning dataset by Tsuchiya \ea~\cite{tsuchiya2025blockchainaddresspoisoning}. As we mentioned before, the dataset provided by Tsuchiya \ea unfortunately has a gap between blocks 20,208,026 and 21,565,106.  Furthermore, by matching transactions based on their \texttt{from} and \texttt{to} addresses, we identify a total of 3,402,101 transactions involving address pairs previously associated with address poisoning attacks—nearly twice the number reported in the original dataset. Since these additional transactions were not included in the published dataset, our results suggest that our methodology identifies approximately 50\% more address poisoning transactions than the existing approach. Extending this analysis to our complete measurement period, we identify 3,757,045 state-invariant transactions, accounting for approximately 16\% of all state-invariant transactions on Ethereum.

Of these transactions on Ethereum, 2,777,055 (approximately 72\%) execute successfully. To analyze the targeted addresses, we cluster recipient addresses by the first and last four hexadecimal characters of the addresses appearing in the emitted \texttt{Transfer} events. We find that more than 90\% of these clusters occur fewer than ten times, with a median cluster size of two, indicating that address poisoning attacks typically target newly generated look-alike addresses. Nevertheless, a small number of large clusters exist. The largest cluster, corresponding to the pattern ``\texttt{0xa9d1$\cdots$3e43}'', contains approximately 151,248 transfer events.

Since the dataset of Tsuchiya \ea~\cite{tsuchiya2025blockchainaddresspoisoning} is limited to Ethereum, we cannot perform the same direct comparison on Optimism and Base. Instead, we investigate whether attacker addresses identified on Ethereum also appear in state-invariant transactions on these Layer-2 networks. We observe similar patterns on both chains, identifying 278,760 such transactions on Optimism and 278,25 on Base. These counts are substantially lower than on Ethereum. One possible explanation is that because our analysis relies on wallet addresses identified from the Ethereum address poisoning dataset from Tsuchiya \ea, these same wallets may have limited activity on Base and Optimism, reducing the coverage of poisoning-related transactions on those networks.

\subsection{On-Chain Postings and Inscriptions}

we identify 39,879 inscription transactions on Ethereum, 603 on Optimism, and 4,470 on Base accounting only for 0.2\% on Ethereum, and less than 0.001\% on Base and Optimism. This corresponds to 
The corresponding median calldata sizes are 148 bytes on Ethereum, 116 bytes on Optimism, and 30 bytes on Base. Compared to the surge observed in September 2023~\cite{messiasinscription}, these results indicate that inscription activity has declined substantially across all three networks.

Overall, inscriptions account for only a negligible fraction of the state-invariant transactions identified in our dataset. Although inscriptions consume blockspace without modifying the blockchain state beyond the payment of transaction fees, their prevalence is several orders of magnitude lower than that of speculative MEV transactions. These findings suggest that, despite the considerable attention inscriptions have received as a source of blockchain bloat, they are not a major contributor to the state-invariant transaction volume observed on Ethereum or the examined Layer-2 networks.

Another important use case for blockchains is on-chain posting or message passing. Rather than transferring assets, users interact with smart contracts to exchange information or coordinate actions by typically emitting events that can be consumed by off-chain services.
To identify such applications, we examine state-invariant transactions originating from smart contracts that are not labeled as MEV bots and whose source code has been verified on Etherscan, Optimistic Etherscan, or BaseScan. We then manually inspect the verified contracts to determine their functionality.
To only focus on contracts that predominantly generate state-invariant activity, we restrict our analysis to contracts that have produced at least 10,000 successful state-invariant transactions and for which more than 99\% of all successful transactions are classified as state-invariant. This threshold filters out high-volume infrastructure contracts, such as DEX routers, which may generate many state-invariant transactions but whose primary functionality involves state-changing operations.

On Ethereum, we identify two contracts that primarily participate in message passing: \texttt{0x166fd42
99364B21c7567e163d85D78d2fb2f8Ad5} and \texttt{0x231055A0852D67C7107Ad0d0DFeab60278fE6AdC}. Together they account for 43,294 or 0.2\% of  state-invariant transactions. On Optimism, we identify the contract \texttt{0x64812F1212f6276068A0726f4695a6637DA3E4F8}, which is labeled on Optimistic Etherscan as a DMail Executor, a decentralized messaging application, accounting for 376,681 or 0.3\% of state-invariant  transactions. On Base, we identify ten such contracts, which together account for approximately 24 million or 2\% of state-invariant transactions.

\section{Discussion}

Our results show that speculative MEV is a major source of spam transactions on Layer-2 blockchains such as Optimism and Base, but not the only one. We also confirm that these transactions consume a substantial amount of gas and blockspace, reducing the effective throughput available to legitimate users while accelerating the growth of blockchain archive nodes. This is particularly concerning because archive nodes are essential for independently verifying blockchain state and querying historical data. Reliance on centralized infrastructure providers for this information undermines the decentralization guarantees of blockchains and introduces trust assumptions regarding the correctness and availability of the returned data.

At the same time, our findings indicate that speculative MEV is often neither the most efficient nor the most profitable extraction strategy for searchers. Consequently, it imposes costs on all parties involved: blockchain operators incur additional storage and execution overhead, users experience increased contention for blockspace, and searchers frequently pay transaction fees for unsuccessful extraction attempts.

Beyond speculative MEV, we show that existing methodologies substantially underestimate the prevalence of spam transactions, particularly on Ethereum. A significant fraction of state-invariant transactions is instead associated with phishing campaigns, most notably address poisoning attacks. These findings suggest that protocol-level mechanisms capable of filtering transactions that do not produce meaningful state changes could significantly reduce blockchain spam while simultaneously limiting the spread of phishing transactions and slowing blockchain growth.

Such an approach, however, would also affect legitimate applications that intentionally use transaction calldata as a communication or data-publication channel, including on-chain messaging systems and inscriptions. Although these applications leverage blockchains as censorship-resistant storage, whether permanently recording arbitrary data on a globally replicated ledger is desirable remains an open question. Previous work \cite{MatzuttHHZMHW18} has shown that illegal content, including child sexual abuse material, has been embedded in blockchain data, raising legal and ethical concerns for operators maintaining full copies of the blockchain.

\section{Related Work}\label{sec:related_work}

Daian et al.~\cite{daian2020flash} introduced the concept of \gls{MEV} arising from transaction reordering, insertion, and censorship. Subsequent work quantified \gls{MEV} at ecosystem scale~\cite{Qin@SP22,McLaughlin@SEC} and characterized major extraction strategies, including arbitrage, liquidations, sandwich attacks, frontrunning, flash-loan attacks, copycat attacks, private-order-flow extraction, non-atomic arbitrage, cross-chain arbitrage, and CEX--DEX arbitrage~\cite{Qin@FC21,Zhou-hft@SP,torres2021frontrunner,Qin-Immitation@SEC,Weintraub-IMC22,Heimbach-MEV-SP-24,Oz-Cross-chain-25,Wu-AFT25}. Unlike these works, which begin by identifying specific \gls{MEV} strategies, our methodology first identifies state-invariant transactions and subsequently attributes them to \gls{MEV} or non-\gls{MEV} activity.

Rollups fundamentally change \gls{MEV} incentives through centralized sequencing, low transaction fees, and the absence of public mempools. Prior work has characterized \gls{MEV} across Layer-2 networks~\cite{Torres-rollingintheshadows-CCS,gogol2024crossrollupmevnonatomicarbitrage}, studied speculative and revert-based \gls{MEV}~\cite{solmaz@AFT25,gogol2025priorityfailsrevertbasedmev,wu2026wait}, and analyzed spam generated by priority mechanisms such as Timeboost~\cite{messias2025expresslanespamcentralization}. Existing approaches typically detect spam only after identifying arbitrage bots, reverted transactions, or strategy-specific traces. In contrast, our methodology identifies state-invariant transactions independently of their origin and subsequently classifies them into speculative, non-speculative, and hybrid \gls{MEV}.

State-invariant transactions are not limited to \gls{MEV}. Address poisoning attacks generate misleading transaction histories through zero-value transfers~\cite{tsuchiya2025blockchainaddresspoisoning}, while inscriptions use transaction calldata to publish arbitrary data without relying on smart-contract state~\cite{messiasinscription}. Other examples include redundant token approvals and self-transfers. These behaviors motivate a broader notion of spam based on the absence of meaningful state changes rather than on bots or transaction intent.

Prior work has proposed techniques for detecting economically exploitable smart contracts~\cite{babel2023clockwork,babel2023lanturn}, formalizing cross-domain \gls{MEV}~\cite{Obadia-Clockwork-21}, and mitigating extractive behavior through mechanisms such as submarine commitments, encrypted mempools, and other ordering defenses~\cite{Breidenbach-Usenix-18,Heimbach-CCS-22,Arka-usenix-24,Bormet-usenix-25,Yang-DeFi-24}. While these approaches primarily target transaction ordering and information leakage, our work highlights a complementary systems challenge: reducing transactions that consume execution, bandwidth, storage, and blockspace without producing meaningful ledger state.

\section{Conclusion}\label{sec:conclusion}

In this paper, we introduced the concept of \emph{state-invariant transactions}, which we defined as transactions whose execution does not modify the blockchain state beyond the payment of transactions fees. We also presented the first large-sale characterization of their prevalence across Ethereum and popular rollups such as Optimism and Base. Our measurements reveal that while such transactions represent only a small fraction of Ethereum activity, they account for a substantial portion of transaction volume and gas consumption on modern rollups, accounting for more than one third of all transactions on Base. These findings show that a significant fraction of the computational resources consumed by blockchain systems is purely associated to transactions that do not change the global state of such systems beyond transaction fees for block producers.

Our analysis further shows that the majority of state-invariant transactions on rollups originate from speculative MEV strategies enabled by low transaction costs and high throughput. On the other hand, Ethereum shows a more heterogeneous ecosystem including reverted transactions, redundant token operations, and phishing campaigns such as address poisoning. By introducing a scalable methodology for identifying state-invariant transactions and distinguishing speculative from non-speculative MEV, we provide a more comprehensive view of blockchain resource utilization than previous works that focus exclusively on successful arbitrage transactions.

Our results also highlights an important system trade-off. Rollups successfully increase throughput and reduce costs, but these same design choices also lower the cost of generating economically reasonable computationally wasteful transactions. Consequently, improvements in scalability can unintentionally incentivize behaviors that consume bandwidth, execution time, storage, and block space without contributing to a meaningful evolution of the blockchain state.

Finally, we believe that state-invariant transactions constitute an important new metric for evaluating blockchain efficiency and should be considered when designing future transaction fee mechanisms, sequencing policies, and spam mitigation techniques. As blockchain ecosystems continue to scale, reducing unnecessary state-invariant execution represents an opportunity to improve resource utilization and allocation, limit malicious activity, and preserve decentralization without compromising the fundamental transparency of decentralized systems.

%------------------------------------------------------------------------------
%
\begin{acks}
This work is supported by the European Union’s Horizon 2020 research and innovation programme under grant agreement No 952226, project BIG (Enhancing the research and innovation potential of Tecnico through Blockchain technologies and design Innovation for social Good) as well as by national funds through FCT, Fundação para a
Ciência e a Tecnologia, under projects UIDB/500
21/2020 (DOI:10.54499/UIDB/50021/2020) and UIDP/50021/2020 (DOI:10.544
99/UIDP/50021/2020).
\end{acks}

% %

%------------------------------------------------------------------------------
\bibliographystyle{ACM-Reference-Format}
\bibliography{references}

\end{document}